\documentclass[aps,prd,superscriptaddress,twocolumn]{revtex4}

\usepackage{amsfonts}
\usepackage{amsmath}
\usepackage{graphicx}
\usepackage{color}
\usepackage{amssymb}

\usepackage[normalem]{ulem}

\begin{document}

\title{Reversed electromagnetic Vavilov-\v{C}erenkov radiation in naturally existing  magnetoelectric media }

\author{O. J. Franca}
\email{francamentesantiago@ciencias.unam.mx}
\affiliation{Instituto de Ciencias Nucleares, Universidad Nacional Aut\'{o}noma de M\'{e}xico, 04510 M\'{e}xico, CDMX, M\'{e}xico}

\author{L. F. Urrutia}
\email{urrutia@nucleares.unam.mx}
\affiliation{Instituto de Ciencias Nucleares, Universidad Nacional Aut\'{o}noma de M\'{e}xico, 04510 M\'{e}xico, CDMX, M\'{e}xico}

\author{Omar Rodr\'\i guez-Tzompantzi}
\email{omar.rodriguez@correo.nucleares.unam.mx}
\affiliation{Instituto de Ciencias Nucleares, Universidad Nacional Aut\'{o}noma de M\'{e}xico, 04510 M\'{e}xico, CDMX, M\'{e}xico}
\begin{abstract}
We consider two semi-infinite  magnetoelecteric media separated by a planar interface whose electromagnetic response is described by axion electrodynamics. The time-dependent Green's function characterizing this geometry is obtained by a method that can be directly generalized to cylindrical and spherical configurations of two magnetoelectrics separated by an interface. We establish the  far-field approximation of the Green's function and apply these results to the  case of a charged particle  moving from one medium to the other at a  high constant velocity  perpendicular  to the interface. From the resulting  angular distribution of the  radiated energy per unit frequency  we provide   theoretical  evidence for the emergence of  reversed Vavilov-\v{C}erenkov radiation in naturally existing magnetoelectric media.   In the case where one of the 
magnetoelectrics is a 3D topological insulator, ${\rm TlBiSe}_2$,  for example, located in front of a regular insulator, we estimate that an average  forward Vavilov-\v{C}erenkov radiation with  frequency $\sim 2.5 \,\, {\rm eV}$ ($\sim 500\,\, {\rm nm}$) will produce a highly suppressed reversed Vavilov-\v{C}erenkov radiation which can be characterized by an effective frequency  in the range of $\sim (4\times 10^{-3}-0.5) \,\, {\rm meV}$. However, this value compares favorably  with recent measurements in left-handed metamaterials  yielding reversed Vavilov-\v{C}erenkov radiation  with frequencies of the order of $(1.2-3.9)\times 10^{-2}\,\, {\rm meV}$.

\end{abstract}

\maketitle
\section{Introduction}

\label{INTRO}

Since its experimental discovery in $1934$  \cite{cherenkov,Vavilov}, Vavilov-\v{C}erenkov (VC) radiation   has played a special role in the study of high-energy particle physics, high-power microwave sources and nuclear and cosmic-ray physics  \cite{Jelley,Jelley1},  both theoretically and phenomenologically. VC radiation occurs when charged particles travel across a  dielectric medium and propagate through the material with velocity  $v$ higher than $c/\sqrt{\epsilon \mu}$. 
Here $c$ is the speed of light in vacuum, $\epsilon$ is the permittivity of the medium, $\mu$ is its permeability, which we take equal to one, and $n=\sqrt{\epsilon}$ is  the refraction index of the material. Throughout this paper, we use Gaussian units. The first theoretical description of such radiation in the framework of Maxwell's theory, developed by  Frank and Tamm in Ref. \cite{FT}, revealed its unique polarization and directional properties.  In particular, VC radiation is produced in a forward cone 
defined by the  angle $\theta_{C}= \cos^{-1}[c/vn] < \pi/2$ with respect to the direction of the incident charge. Since the emergence of accelerators in nuclear and high-energy physics, VC radiation has been widely used to design an impressive variety of detectors, such as e.g. the ring imaging \v{C}erenkov detectors \cite{Ypsilantis}, which can identify  charged particles by providing  a straightforward effective tool to test its physical properties like velocity, energy, direction of motion and charge \cite{PADG}. As remarkable cases, the antiproton \cite{Chamberlain}  and the J particle \cite{Aubert} were discovered using \v{C}erenkov  detectors.

In recent years, the study of the reversed VC radiation, an exotic electromagnetic phenomenon, has long been of considerable interest \cite{Pendry,Pendry2, LU,  Luo,  RRCR, Sheng, Chen, Duan,Tao}. The reversed VC radiation occurs when the photons are emitted in the backward direction with respect to the velocity of  the propagating charged particle.  This remarkable analytic prediction was introduced by  Veselago in Ref.\cite{Veselago}, invoking materials having simultaneously a negative permittivity and permeability, dubbed as left-handed media (LHM), as opposed to normal dielectrics called 
right-handed media (RHM). This distinction is better characterized in terms  of an electromagnetic wave with electric field ${\mathbf{\bar E}}$,  magnetic field ${\mathbf{\bar  B}}$, and  wave vector ${\mathbf k}$  propagating in a linear medium where  the conditions ${\mathbf k} \cdot {\mathbf{\bar  E}}=0$ and ${\mathbf k}\cdot {\mathbf{ \bar B}}=0$ hold. A medium is said to be left-handed (right-handed) according to ${\mathbf k}$ being  antiparallel (parallel) to ${\mathbf {\bar E}}\times {\mathbf{\bar  B}}$, i. e., when the group velocity and the phase velocity of the wave  are  antiparallel (parallel). The realization of  a negative refraction index by creating an interface between a LHM and an ordinary media was also considered in  Ref. \cite{Veselago} and experimentally proven later in Ref. \cite{SHELBY}.
Since LHM are not readily available in nature, they have been artificially constructed  in the laboratory as metamaterials, e.g., by combining metallic thin wires \cite{Pendry} with split-ring resonators \cite{Pendry2} in a periodic cell structure. Also plasmonic thin films \cite{LU} and photonic crystals  \cite{Luo}  constitute an alternative.

In this paper, we show that reversed VC radiation  occurs in magnetoelectric media  \cite{ODELL},  which are  naturally existing RHM like {antiferromagnets \cite{Hehl}},  topological insulators (TIs) \cite{TI1,TI11,QILIBRO}, and  Weyl semimetals \cite{ARM}, for example.  It is interesting to observe that Reversed \v{C}erenkov sound in topological insulators has been  already discussed in Ref. \cite{SMIRNOV}.
Linear magnetoelectric media are characterized  by the magnetoelectric polarizability  tensor $\alpha_{ij}$, which is defined as the response of the  magnetization under a change of an electric field or, equivalently, as the response of the  polarization under a change of a magnetic field, i.e.,
\begin{equation}
\alpha_{ij}=\left[ \frac{\partial M_j}{\partial E_i}\right]_{{\mathbf B}={\mathbf 0}}=\left[ \frac{\partial P_i}{\partial B_j}\right]_{{\mathbf E}={\mathbf 0}} \equiv
 {\tilde \alpha}_{ij}+ \frac{e^2}{\hbar c}\frac{\vartheta}{4 \pi^2 } \delta_{ij},\label{MEPOL}
\end{equation}
where $\vartheta$ is a dimensionless parameter. The equality follows because $M_j=\partial {\mathcal E}/\partial B_j$ and  $P_i=\partial {\mathcal E}/\partial E_i$, with  ${\mathcal E}$ being the electromagnetic enthalpy density \cite{VDBPRL102_2009}. Here ${\tilde \alpha}_{ij}$ denotes the traceless part of $\alpha_{ij}$ . Leaving aside the remarkable  microscopic properties of magnetoelectric media, we will consider only  the effective-medium theory describing their electromagnetic response  arising from the term proportional to $\delta_{ij}$ in Eq. (\ref{MEPOL}), i. e., assuming an isotropic material. Such effective description   is provided by the standard Maxwell Lagrangian ${\cal L}_{\rm ED}$  plus the additional term ${\cal L}_\vartheta = -\frac{e^2}{\hbar c}\frac{\vartheta}{4 \pi^2 }\,  \mathbf{E}\cdot \mathbf{B}$. Here  $\mathbf{E}$ and $\mathbf{B}$ are the electromagnetic fields, $\alpha$ is the fine-structure constant and  $\vartheta$ is a field known as the (scalar) magnetoelectric polarizability  (MEP) in condensed matter physics or the axion field in particle physics \cite{axion}. In this way,  the addition  of ${\cal L}_\vartheta$  to the Maxwell Lagrangian is usually referred to as the Lagrangian density for  axion electrodynamics. However, we  consider $\vartheta$ as an additional parameter characterizing the material, in the same footing as its permittivity $\epsilon$ and permeability $\mu$, which lead us to restrict the designation of axion electrodynamics to $\vartheta$ electrodynamics ($\vartheta$-ED) emphasizing  that $\vartheta$ is not a dynamical field. The nature of the MEP depends on the type of magnetoelectric material under consideration and it is deeply related to the magnetic symmetries of the substance \cite{DZYA,Rivera}  and/or to the  properties  of its band structure \cite{TI11}. It can be calculated  from a Kubo-type response formula, once a microscopic model Hamiltonian for the material is adopted. The permittivity $\epsilon$ is usually designed  by the Drude-Lorentz type of single resonance oscillator  model \cite{Mohideen}.  As as first step 
in dealing with radiation,  we consider $\vartheta$ and  $\epsilon$ frequency independent. 

The main feature characterizing magnetoelectric materials, which is responsible for most of their unusual effects, is the magnetoelectric 
(ME) effect  arising from the additional contribution ${\cal L}_\vartheta$ \cite{QI_SCIENCE}. This coupling produces effective field-dependent  charges and current densities, which allow the generation  of an electric (magnetic) polarization due to the presence of a magnetic (electric) field.  Let us emphasize that we are dealing with standard electrodynamics of a RHM ($\epsilon >1,\, \mu>1,\, n>1$) supplemented only by additional field-dependent sources.

Among the choices of accessible magnetoelectric media, we consider  materials characterized by a  piecewise constant value of the MEP $\vartheta$ and focus on (TIs).  
Three-dimensional (3D) strong TIs are a fascinating class of RHM that can host a  conducting helical surface state with an odd number of fermions at the low-energy limit,  each  having  
the dispersion relation  of a nondegenerate Dirac cone with  a crossing point at or close to the Fermi level $E_{\rm F}$. These gapless crossing points are called Dirac points. Nevertheless, TIs behave as magnetoelectric insulators in the bulk with a finite energy  gap.
 The  surface state is further topologically  protected by time-reversal symmetry and/or inversion symmetry, coupled with spin-momentum locking properties. The latter means that the spin orientation of the  electrons on the surface Dirac cone is always locked perpendicularly to their momentum. 
A distinguishing feature of 3D strong TIs among magnetoelectrics, is that the   dimensionless parameter $\vartheta$, dubbed as the orbital magnetoelectric polarizability in this case, is of topological nature and arises  from the bulk band structure. It is  given by a non-Abelian  Berry flux over the Brillouin zone and results in an integer  multiple of $\pi$ \cite{TI11, VDBPRL102_2009,QZh}.

Some general comments  regarding the properties of the additional term ${\cal L}_\vartheta $ in the case of TIs are now in order. Let us recall that the electric and magnetic fields have dimensions of charge divided by distance squared in Gaussian units (for the moment we retain $ \hbar$ and $c$). In this way, the contribution of the Lagrangian density ${\cal L}_\vartheta$ to the action is $S_\vartheta=(c \hbar^2/e^4)\int dt \, d^3x \, {\cal L}_\vartheta $. Rewriting ${\mathbf E}\cdot {\mathbf B} $ in terms of the field strength tensor  $F_{\alpha\beta}=\partial_\alpha A_\beta-\partial_\beta A_\alpha$, where   $A_\mu$ is the electromagnetic potential, and considering a  
closed spacetime with no boundaries, we get 
\begin{equation}
\frac{S_\vartheta}{\hbar}=\frac{\vartheta}{32 \pi^2 }\int d^4x \epsilon^{\alpha\beta\mu\nu} \frac{1}{e^2} F_{\alpha\beta} F_{\mu\nu}=  \vartheta\,  C_2, \label{THETAACT}
\end{equation}
where $C_2$ is an integer. This is because in such spaces the dimensionless integral in Eq. (\ref{THETAACT}) is equal to $32 \pi^2\, C_2$, where $C_2$  is the second Chern number of the manifold \cite{FUJIKAWA}. Under changes of $\vartheta$, the quantity $\exp(-i S_\vartheta/\hbar)$ must remain invariant, which means that two values of $\vartheta$ differing by an integer  multiple of $2 \pi$ are equivalent. Further imposing time-reversal (TR) symmetry  yields new constraints  on the values of $\vartheta$. Since $ \mathbf{E}\cdot \mathbf{B}$ is odd under TR, one could think that the only allowed value would be $\vartheta=0$ (modulo  $2 \pi)$. Nevertheless, the condition $\exp(-iS_\vartheta/\hbar)= \exp(+iS_\vartheta/\hbar)$ yields the possibility of having $\vartheta= \pi $. In this way, we obtain two families of magnetoelectric materials described by the choices $\vartheta_1= 0$ (normal insulators) and $\vartheta_2= \pi$ (topological insulators). Both values of $\vartheta$ are defined modulo  $2 \pi$.
 
Another important property of $S_\vartheta$ is that the integrand in Eq. (\ref{THETAACT}) is  a total derivative, as can be readily seen by recalling the Bianchi identity $\epsilon^{\alpha\beta\mu\nu}\partial_\beta F_{\mu\nu}=0$  . In this way, modifications to Maxwell equations only arise when $\partial_\mu \vartheta \neq 0$, as it happens  at the interface $\Sigma $ of two materials having  different constant values of $\vartheta$, for example. In this case, the action (\ref{THETAACT}) can be integrated yielding a $2+1$ action at the boundaries  corresponding to the  Chern-Simons term. This means that at the boundaries of a 3D TI we have a quantum  anomalous Hall (QAH) effect  associated with each value of $\vartheta$, with Hall conductivity given by $\sigma_{{\rm H}}=\vartheta e^2/2 \pi h $. 
In this way, the contribution  to the  total Hall conductivity from the interface $\Sigma$ between  a TI and a regular insulator  is
\begin{equation}
\sigma_H^{\Sigma}=\frac{e^2}{h}\left( \left|\frac{1}{2}\right|+ m\right),
\end{equation} 
since two values of $\vartheta$ differing by an integer multiple $m$ of $2\pi$ are  equivalent.

The half integer contribution to $\sigma^\Sigma_H$ is a bulk property, which allows us to distinguish this case from that of a 2D surface gapped crystal having $\sigma_H= N e^2/h$, with $N$ integer, thus showing that both conditions  are not topologically equivalent. When dealing with a TR-invariant TI  in a region with no boundaries, the number $m$ remains undetermined. The integer part of $\sigma^\Sigma_H$
becomes resolved   only in the presence of a boundary between two TIs with different values of $\vartheta$, when  TR symmetry is broken  by  gapping the interface.  In this  way we provide an adiabatic transition between those two topologically inequivalent insulators and the value of $m$  depends on  the specific properties of such breaking. Such TR symmetry breaking is usually realized  by an external magnetic field across the interface  or by  a  magnetic doping of the surface.

The existence of TIs was predicted in Refs.  \cite{Kane-Mele
1,Kane-Mele 2, Bernevig-Hughes-Zhang} and their observation in two-dimensional HgTe/CdTe quantum wells was reported in  Ref. \cite{Koenig et al}. Then the authors in Refs.  \cite{Fu-Kane-Mele, Moore-Balents, Roy} discovered that the
topological characterization of the quantum spin Hall insulator state has a
natural generalization in three dimensions. Shortly afterward, this behavior
was predicted in several real materials \cite{Fu-Kane} which included Bi$%
{}_{1-x}$Sb$^{}_{x}$ as well as strained HgTe and $\alpha$-Sn. Subsequently,  the experimental discovery of the first three-dimensional TIs in Bi${}_{1-x}$Sb$^{}_{x}$ was reported in Ref.
\cite{Hsieh et al}. Later, a second generation of
TIs, such as Bi${}_{2}$Se${}_{3}$, Bi${}_{2}$Te${}_{3}$, and Sb${}_{2}$Te$%
{}_{3}$, was identified theoretically in Ref. \cite{Xia et al} and experimentally  discovered in Refs. 
\cite{Xia et al, Zhang et al}. This led  to the detection of a huge variety of TIs  \cite{ANDO} and to their subsequent classification in  a periodic table where different classes of these materials can be identified, distinguishing them
by the presence or
absence of time-reversal, particle-hole, and chiral symmetry \cite{TI11}. 
A new class of TIs, called axion insulators (AXIs), has been recently
proposed as a new arena to probe topological phases. They have the same bulk MEP $\vartheta=\pi$  as 3D TIs, with gapped both bulk and surfaces, where the topological index is protected by inversion symmetry, instead of time-reversal symmetry. They are expected to show up in magnetically doped TI heterostructures  with magnetization pointing inward and outward from the top and bottom interfaces of the TI \cite{DiXiao,Mogi1}. Also, the possibility of having a different class of  intrinsic axion insulators, which do not require magnetic doping,  has been investigated in a pyrochlore lattice in Ref. \cite{Varnava}.

When time reversal is broken at the interface between a  3D strong TI and a regular one, either by the application of a magnetic coating and/or by doping the TI with transition metal elements, the opening of the gap in the surface states induces  several exotic phenomena that can be tested experimentally. Among them we find  the (QAH) effect, the quantized magneto-optical effect,   
topological magnetoelectric  (TM) effect, and the image magnetic monopole effect, all of which are a direct consequence of the change in the parameter $\vartheta$ between the two phases.
These effects can be also produced in AXIs.

The QAH effect has already been experimentally observed in thin films of the magnetic chromium-doped topological insulator 
$({\rm Bi,Sb})_2{\rm Te}_3$ \cite{Chang 1}. In Ref. \cite{Kou} the observation of the QAH effect in extremely thin films of  the magnetic topological insulator (Cr$_{0.12}$Bi$_{0.26}$Sb$_{0.62}$Sb$_{0.62}$)$_2$Te$_3$ is reported.  The characteristic behavior of the QAH effect has also been shown experimentally in thin films of the topological insulator  Cr$_x$(Bi$_{1-y}$Sb$_y$)$_{2-x}$Te$_3$, which were grown on semi-insulating InP(111) substrates using molecular beam epitaxy methods \cite{Checkelsky}. Also, Ref. \cite{Chang 2} demonstrates a high-precision confirmation of the QAH state in V-doped (Bi,Sb)$_2$Te$_3$ films, which is a strong ferromagnetic TI. The  QAH effect has also been observed in AXIs \cite{Mogi}.
Employing  terahertz time-domain spectroscopy, quantized magneto-optical effects have been observed  by measuring the  topological Faraday and Kerr rotations in QAH states on  Cr$_x$(Bi$_{0.26}$Sb$_{0.74}$)$_{2-x}$Te$_3$ magnetic TI films. In this work the authors also report the observation of the QAH together with an experimental indication of the TM effect \cite{Okada}.  Quantized Faraday and Kerr rotations in magnetic fields higher than 5 T  were observed in the 3D topological insulator  Bi$_2$Se$_3$, providing  evidence of the TM effect by an indirect measurement of the value $\vartheta=\pi$ \cite{Wu}. Reference \cite{Dziom}  reported a quantized Faraday rotation in high external magnetic fields when linearly polarized terahertz radiation  passes through the two surfaces of a strained HgTe 3D TI. This constitutes a direct consequence of the TM effect, thus  confirming  axion electrodynamics  as the effective theory describing the response of   3D strong TIs.

Our following discussion of the reversed VC radiation parallels the analysis of VC radiation in standard references \cite{Schwinger,Jackson, Panofsky}. We deal with radiation by further extending to the time-dependent case the Green's function method already developed in Refs. \cite{UrrutiaMartinCambiaso3,UrrutiaMartinCambiaso} to study the static electromagnetic response of TIs.

The paper is organized  as follows. In Sec. \ref{theta-ED}, we review the effective theory describing the electromagnetic response of magnetoelectrics, to be called  $\vartheta $-ED. Section \ref{GF method} contains the  calculation of  the time-dependent Green's function (GF) for the electromagnetic potential $A_\mu$ arising from arbitrary sources in the presence of two semi-infinite magnetoelectric media with different values of the parameter $\vartheta $, each having the same permittivity $\epsilon$, and separated by a planar interface that encodes the manifestation of the ME effect. 
This  section also reports the results of the far-field approximation of the GF, which is required to deal with radiation. We give an analytic expression of  the GF in this limit, whose detailed calculations are presented in the  Appendix \ref{GFfarzone}. In this way, we determine  how the phase of the GF in standard ED \cite{Schwinger} changes due to the presence of the magnetoelectric media, thus explaining the origin of reversed VC radiation. 
In Sec. \ref{RCRSEC}, we consider  
a charged particle moving with constant velocity $v > c/\sqrt{\epsilon}$ in the direction perpendicular to the interface between the two magnetoelectrics. Our formalism yields analytic results for the physical quantities involved. The far-field expressions for ${\mathbf E}$ and ${\mathbf B}$ are calculated, together with  the angular distribution of the total radiated energy per unit  frequency  $d^2 E/d\omega d\Omega$. The resulting  angular distribution indicates the presence of reversed VC radiation, which is the most important conclusion of our work. Its presence  is further illustrated in polar plots showing the angular distribution of the reversed VC radiation together with those corresponding to the standard forward VC  radiation. 
In Sec. \ref{Total E}, we calculate  the total radiated energy per unit frequency $d E/d\omega$, both in  the forward and the backward directions, together with  the number of photons per unit length radiated in the backward direction. The power per unit frequency radiated in the backward direction is also  obtained. Section  \ref{NUM_EST} is devoted to some  numerical estimations of our results  considering the topological insulator ${\rm TlBiSe}_2 $  as one of the semi-infinite media and a regular insulator with the same permittivity as the other. We have chosen this material  because its electric and topological properties are known from previous references. We emphasize that we are  not doing any {\it ab initio} or experimental calculation,  either on this material or in any other sector of the manuscript. Let us insist  again that our approach is based on the effective electromagnetic response of magnetoelectrics encoded in $\vartheta$-ED. From this point of view, we assume that any material we refer to 
has been fully characterized, meaning that their macroscopic parameters have been already determined. Section \ref{Summary} comprises a concluding summary of our results. In the following, we set  $\hbar=c=1$, we denote by $\eta^\mu{}_\nu$ the Minkowski metric with signature  $(+,-,-,-)$, and we take the convention $\varepsilon^{0123}=1$ for the Levi-Civita symbol.

\section{$\vartheta $ Electrodynamics}
\label{theta-ED}

\begin{figure}[tbp]
\centering 
\includegraphics[width=8CM]{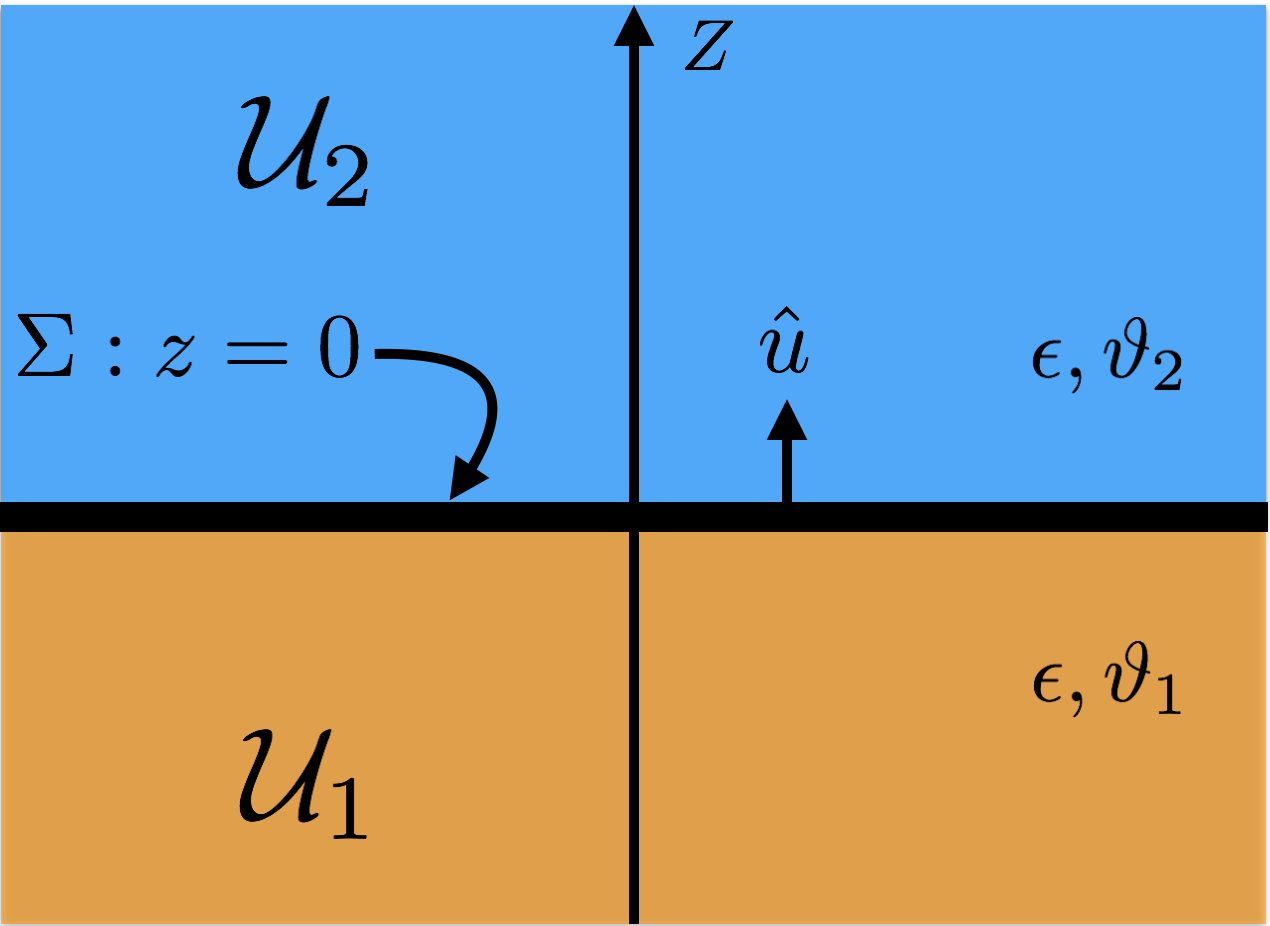} 
\caption{Region over which $\vartheta$-ED is defined, with a
uniform dielectric constant $\epsilon $.}
\label{REGIONS}
\end{figure}

Let us consider two semi-infinite magnetoelectric media separated by a planar interface $\Sigma$ located at $z=0$, filling the regions $\mathcal{U}_{1}$ and $\;\mathcal{U}_{2}$ of the space, as shown in Fig. (\ref{REGIONS}).  Motivated by the results in Ref. \cite{UrrutiaMartin}, whereby the effects of the $\vartheta $-term are substantially enhanced with respect to the optical contributions when both $\vartheta $ media have the same permittivity, we take $\epsilon _{1}=\epsilon _{2}=\epsilon $. In our case, we want to suppress the transition radiation, which occurs whenever a charge propagates across two media with different permittivities \cite{TRANSITION}. This radiation would be  present for all particle  velocities and would interfere with the VC radiation we are interested in, making unnecessarily  complicated its
 theoretical discussion, together with hindering its possible experimental verification. Additionally, we assume that the parameter $\vartheta $ is piecewise constant taking the  values $\vartheta =\vartheta _{1}$ in the region $\mathcal{U}_{1} $ and  $\vartheta =\vartheta _{2}$ in the region $\mathcal{U}_{2}$. This is expressed as 
\begin{equation}
\vartheta (z)=H(z)\vartheta _{2}+H(-z)\vartheta _{1}, \end{equation}
where $H(z)$ is the Heaviside function with $H(z)=1,\;\;z\geq 0\;$, and $H(z)=0,\;\;z<0.$
In this restricted case we can take the action for  the effective theory describing the electromagnetic response of this media as 
\begin{eqnarray}
S[\Phi ,\mathbf{A}]&=&\int dt\,d^{3}\mathbf{x}\left[ \frac{1}{8\pi }\left(
\epsilon \mathbf{E}^{2}-\frac{1}{\mu }\mathbf{B}^{2}\right)\right.\nonumber\\
&&\left. -\frac{e^2}{4\pi ^{2}}\vartheta (z)\,\mathbf{E}\cdot \mathbf{B}-\varrho \Phi +%
\mathbf{J}\cdot \mathbf{A}\right],  \label{Action1}
\end{eqnarray}
where $\varrho$ ($%
\mathbf{J}$) are  charge (current) densities  and  $e^2=\alpha$ is the fine-structure constant.

Let us  emphasize that we are describing each medium  by a constant MEP $\vartheta $ in the bulk,  which has the value 
$\vartheta=\pi$ ($\vartheta=0$) for  a topological insulator  (regular insulator). While there are  boundary effects arising from the gapping of the interface, the  MEP  $\vartheta=\pi$ of the  bulk of the 3D topological insulator has a  gauge-invariant and topological origin.

The electromagnetic
fields $\mathbf{E}$ and $\mathbf{B}$ are related  with the electromagnetic
potentials $\Phi $\ and $\mathbf{A}$\ in the standard form 
\begin{equation}
\mathbf{E=-}\frac{\partial \mathbf{A}}{\partial t}-\nabla \Phi ,\;\;\;\;\;%
\mathbf{B=\nabla \times A},  \label{DEF_POT}
\end{equation}%
as a consequence of the homogeneous
Maxwell equations%
\begin{equation}
\nabla \cdot \mathbf{B}=0,\;\;\;\;\;\nabla \times \mathbf{E}=-\frac{\partial 
\mathbf{B}}{\partial t}.  \label{HOMEQ}
\end{equation}%

\

It can be shown that the term ${\mathbf E}\cdot {\mathbf B} $ is a total derivative, which tells us that there are no modifications to the dynamics in the bulk. In this way, all the effects induced by ${\cal L}_\vartheta $ arise on  the interface and will  manifest themselves as a consequence of the boundary conditions there. Performing the
variation of the action (\ref{Action1}) gives the following set of modified Maxwell
equations:
\begin{eqnarray}
\epsilon \nabla \cdot \mathbf{E} &=&4\pi \varrho +\tilde{\theta}\delta (z)%
\mathbf{B}\cdot \mathbf{\hat{u}}\;,  \label{Gauss E} \\
\nabla \times \mathbf{B}-\epsilon \frac{\partial \mathbf{E}}{\partial t}
&=&4\pi \mathbf{J}+\tilde{\theta}\delta (z)\mathbf{E}\times \mathbf{\hat{u}}%
\;.  \label{Ampere}
\end{eqnarray}%
where $\mathbf{\hat{u}}$ is the outward unit vector normal to the region ${\cal U}_1$ and 
\begin{equation}
{\tilde{\theta}=\alpha (\vartheta _{2}-\vartheta _{1})/\pi.}
\label{TILDE_THETA}
\end{equation}
In the case of a TI located in region $2$  of Fig. \ref{REGIONS} (${\vartheta}_2=\pi$), in front of a regular insulator ($\vartheta=0$) in region $1$,  we have 
\begin{equation}
{\tilde \theta}=\alpha(2m+1), \label{TILDE_THETA_1}
\end{equation}
with $m$ being an  integer depending on the details of the TR symmetry breaking at the interface.
For
definiteness, we deal with media having $\mu =1$.  The novelty in the
above equations is that they introduce additional field-dependent effective charge and
current densities%
\begin{equation}
\varrho _{\vartheta }=\frac{1}{4\pi }\tilde{\theta}\delta (z)\mathbf{B}\cdot 
\mathbf{\hat{u}},\qquad \mathbf{J}_{\vartheta }=\frac{1}{4\pi }\tilde{\theta}%
\delta (z)\mathbf{E}\times \mathbf{\hat{u}},  \label{EFFSOURCES}
\end{equation}%
with support only on the interface $\Sigma$  between the two media. Consequently, the
standard Maxwell equations remain valid in the bulk. The conservation equation 
\begin{equation}
\nabla\cdot{\mathbf J}_\vartheta + \frac{\partial \varrho_\vartheta}{\partial t}=0
\end{equation}
can be readily verified  by  using  
Faraday's law together with the relation 
\begin{equation}
\left[ \partial _{i}\delta (z)\right] \varepsilon _{ijk}E_{j}u_{k}=\delta
^{\prime }(z)\delta _{i}^{3}\varepsilon _{ijk}E_{j}u_{k}=0,
\end{equation}%
which arises since  the only nonzero component  of $u_{k}$  is $u_{3}=1$.  

 The terms proportional to $\tilde \theta$ in Eqs. 
(\ref{Gauss E}) and  (\ref{Ampere}) describe the
ME effect which is the distinctive feature of  $\vartheta $-ED.
Let us remark that Eqs. (\ref{Gauss E}) and (\ref{Ampere}) can also be obtained starting  from the standard Maxwell equation in a  material medium \cite{Schwinger,Jackson},
\begin{eqnarray}
\nabla \cdot \mathbf{D}=4\pi \varrho &,&\;\nabla \times \mathbf{H}=\frac{%
\partial \mathbf{D}}{\partial t}+4\pi \mathbf{J}\;, \\
\nabla \cdot \mathbf{B}=0 &,&\;\;\;\nabla \times \mathbf{E}=-\frac{\partial 
\mathbf{B}}{\partial t}\;,
\end{eqnarray}%
with the constitutive relations 
\begin{equation}
\mathbf{D}=\epsilon \mathbf{E}-\frac{\alpha }{\pi }\vartheta(z)\mathbf{B},\quad 
\mathbf{H}=\mathbf{B}+\frac{\alpha }{\pi }\vartheta(z)\mathbf{E}.
\end{equation}

Assuming that the time derivatives of the fields are finite in the vicinity
of the interface, the modified Maxwell equations (\ref{Gauss E})
and (\ref{Ampere}) yield the following boundary conditions (BCs) 
\begin{eqnarray}
&&\hspace{-0.5cm}\epsilon \left[ \mathbf{E}_{z}\right] _{z=0^{-}}^{z=0^{+}}=\tilde{\theta}%
\mathbf{B}_{z}|_{z=0}, \quad    \left[{\hat {\mathbf u}} \times \mathbf{B}\right] _{z=0^{-}}^{z=0^{+}}=
-\tilde{\theta}({\hat {\mathbf u}} \times \mathbf{E})|_{z=0}\;,  \label{Conditions 1} \nonumber \\
&&  \\
&&  \hspace{-0.5cm} \left[ \mathbf{B}_{z}\right] _{z=0^{-}}^{z=0^{+}}=0, \quad \left[ {\hat {\mathbf u}} \times \mathbf{E})\right] _{z=0^{-}}^{z=0^{+}}=0\;,  \label{Conditions 2}
\end{eqnarray}%
for vanishing external sources at $z=0$. These BCs are derived
either by
integrating the field equations over a pill-shaped region across the interface or by using the Stokes theorem over a closed rectangular circuit perpendicular to the interface \cite{Schwinger,Jackson}.
The notation is $\left[ \mathbf{V}\right] _{z=0^{-}}^{z=0^{+}}=\mathbf{V}%
(z=0^{+})-\mathbf{V}(z=0^{-})$, $\mathbf{V}\big|_{z=0}=\mathbf{V}(z=0)$,
where $z=0^{\pm }$ indicates the limits $z=0\pm \eta $, with$\;\eta
\rightarrow 0$,$\;$ respectively.  Here the subindex $\parallel$ denotes the projection of a vector into the $x-y$ plane. The continuity conditions (\ref{Conditions
2}) imply that the right-hand sides of the discontinuity conditions (\ref%
{Conditions 1}) are well defined and they represent self-induced surface
charge and surface current densities, respectively. The BCs (\ref{Conditions 1}) clearly illuminate again  the ME effect, which is localized
just at the interface $\Sigma$  between the two media.

\section{Green's function method}

\label{GF method}

In this section, we extend to the case of time-dependent $\vartheta $-ED with planar symmetry the Green's function method discussed in Ref. \cite{UrrutiaMartinCambiaso} for the static limit. A first step in this direction was presented in Ref.\cite{UrrutiaMartinCambiaso3}. The knowledge of the GF allows us to compute the electromagnetic fields for an arbitrary distribution of sources, as well as to solve problems with given Dirichlet, Neumann, or Robin boundary conditions on arbitrary surfaces. In what follows, we restrict ourselves to contributions of free external sources $J^\mu=(\varrho, \mathbf{J})$ located outside the interface $\Sigma$ and  to systems without boundary conditions imposed on additional surfaces, except for the boundary conditions at infinity.

 A compact formulation of the problem can be given in terms of the potential $A^{\mu }=(\Phi ,\mathbf{A})$, which satisfies the equations
\begin{equation}
\left[ \left[ \Box ^{2}\right] _{\;\;\nu }^{\mu }-\tilde{\theta}\delta
(z)\varepsilon _{\quad \;\nu }^{3\mu \alpha }\partial _{\alpha }\right]
A^{\nu }=4\pi J^{\mu },  \label{EQ_POT}
\end{equation}%
in the modified Lorenz gauge 
\begin{equation}\epsilon \frac{\partial \Phi }{\partial t}+\nabla \cdot \mathbf{A}=0,
\end{equation}
 with the operator 
\begin{equation}
\left[ \Box ^{2}\right]_{\;\;\nu }^{\mu }=(\epsilon \Box ^{2},\;\Box ^{2}\delta _{\;j}^{i}), \quad \Box ^{2}=\epsilon \partial _{t}^{2}-\nabla ^{2}.\label{17}
\end{equation}
The current is $J^\mu=(\varrho, {\mathbf J})$.
The BCs (\ref{Conditions 1}) and (\ref{Conditions 2}) reduce to 
\begin{eqnarray}
A^{\mu }|_{z=0^{-}}^{z=0^{+}}=0&,&\epsilon \partial_{z}A^{0}|_{z=0^{-}}^{z=0^{+}}=-\tilde{\theta}\varepsilon _{\quad \;\nu
}^{30\alpha }\partial _{\alpha }A^{\nu }|_{z=0},\nonumber\\
\partial_{z}A^{i}|_{z=0^{-}}^{z=0^{+}}&=&-\tilde{\theta}\varepsilon _{\quad \;\nu
}^{3i\alpha }\partial _{\alpha }A^{\nu }|_{z=0},  \label{BCFG}
\end{eqnarray}
in terms of the vector potential.
Next we introduce the Green's
function $G_{\;\;\sigma }^{\nu }(x,x^{\prime })$\ satisfying 
\begin{equation}
\left[ \left[ \Box ^{2}\right] _{\;\;\nu }^{\mu }-\tilde{\theta}\delta
(z)\varepsilon _{\quad \;\nu }^{3\mu \alpha }\partial _{\alpha }\right]
G_{\;\;\sigma }^{\nu }(x,x^{\prime })=4\pi \eta _{\;\;\sigma }^{\mu }\delta^4
(x-x^{\prime }),  \label{Eq GF temporal}
\end{equation}%
together with the BCs arising from Eq. (\ref{BCFG}), in such a way that the
four potential is 
\begin{equation}
A^{\mu }(x)=\int d^{4}x^{\prime }G_{\;\,\nu }^{\mu }(x,x^{\prime })J^{\nu
}(x^{\prime }),  \label{A and GF}
\end{equation}%
determined up to homogeneous solutions of Eq. (\ref{EQ_POT}). We solve Eq. (\ref{Eq GF
temporal}) along the same lines introduced in Refs. \cite%
{UrrutiaMartinCambiaso,UrrutiaMartinCambiaso3}. First we   take advantage of the translational
invariance in time and in the transverse directions $x$\ and $y$ by introducing  the reduced Green's function $g_{\;\;\nu }^{\mu }(z,z^{\prime };\mathbf{k_{\perp }},\omega )$, such that  \cite{Schwinger}
\begin{eqnarray}
G_{\;\;\nu }^{\mu }(x,x^{\prime })&=&4\pi \int \frac{d^{2}\mathbf{k_{\perp }}%
d\omega }{(2\pi )^{3}}e^{i\mathbf{k_{\perp }}\cdot \mathbf{R}_{\perp
}}e^{-i\omega (t-t^{\prime })}\nonumber\\
&&\times g_{\;\;\nu }^{\mu }(z,z^{\prime };\mathbf{k_{\perp }},\omega ),\nonumber\\
&\equiv& \int_{-\infty }^{\infty }d\omega \;G_{\;\;\nu}^{\mu }({\mathbf x},{\mathbf x}^{\prime };\omega )e^{-i\omega (t-t^{\prime })},
\label{GF temporal coord}
\end{eqnarray}
where $\mathbf{R}_{\perp }=(\mathbf{x}-\mathbf{x^{\prime }})_{\perp}=(x-x^{\prime },y-y^{\prime })$ and $\mathbf{k_{\perp }}=(k_{x},k_{y})$ is the momentum perpendicular to the vector ${\mathbf {\hat u}}$ in Fig. \ref{REGIONS}. The resulting equation  for $g_{\;\;\nu }^{\mu }$ is 
\begin{equation}
\left[ {\mathcal O}^\mu{}_\nu +i\tilde{\theta}\delta
(z)\varepsilon _{\quad \;\;\;\nu }^{3\mu \alpha }k_{\alpha }\right]
g_{\;\;\sigma }^{\nu }(z,z^{\prime };\mathbf{k}_{\perp },\omega )=\eta
_{\;\;\sigma }^{\mu }\delta (z-z^{\prime })\;,  \label{Eq GF temporal Red 1}
\end{equation}%
where \ $k^{\alpha }=(\omega ,\mathbf{k}_{\perp },0)$, and ${\mathcal O}^\mu{}_\nu $ is given by Eq. (\ref{17}) after  the replacements $\nabla^2 \rightarrow -{\mathbf k}^2_{\perp} +\partial^2/\partial z^2$ and $\partial_t \rightarrow -i \omega$ are made.

To solve Eq. (\ref{Eq GF temporal Red 1}) we employ the --- by now --- standard
method \cite{Grosche} to deal with $\delta $-like interactions, whereby a convenient free
GF can be used to integrate the GF equation (\ref{Eq GF temporal Red 1}). By free, here we mean  a GF satisfying Eq. (\ref{Eq GF temporal})\
with $\tilde{\theta}=0$ and which can be written in the same form as in Eq.  (\ref{GF
temporal coord}). To this end, we consider the reduced free Green's function ${\mathcal H}^\rho{}_\nu$
associated with the operator ${\mathcal O}^\mu{}_\nu$
previously defined, which solves 
\begin{equation}
{\mathcal O}^\mu{}_\rho {\mathcal{H}}_{\;\;\nu }^{\rho
}(z,z^{\prime };\mathbf{k}_{\perp },\omega )=\eta _{\;\;\nu }^{\mu }\delta
(z-z^{\prime })  \label{Eq GF temporal Red 2}
\end{equation}%
and satisfies standard boundary conditions at infinity.  From now on, calligraphic letters will  denote free GFs, i.e., GFs that are independent of $\vartheta$. Separating the
components, we have 
\begin{eqnarray}
\epsilon \left( \mathbf{k}_{\perp }^{2}-\omega ^{2}\epsilon -\partial
_{z}^{2}\right) {\mathcal{H}}_{\;\;0}^{0}(z,z^{\prime };\mathbf{k}_{\perp
},\omega ) &=&\delta (z-z^{\prime }),  \notag \\
\epsilon \left( \mathbf{k}_{\perp }^{2}-\omega ^{2}\epsilon -\partial
_{z}^{2}\right) {\mathcal{H}}_{\;\;i}^{0}(z,z^{\prime };\mathbf{k}_{\perp
},\omega ) &=&0,  \notag \\
(\mathbf{k}_{\perp }^{2}-\omega ^{2}\epsilon -\partial _{z}^{2}){\mathcal{H}}%
_{\;\;0}^{i}(z,z^{\prime };\mathbf{k}_{\perp },\omega ) &=&0,  \notag \\
(\mathbf{k}_{\perp }^{2}-\omega ^{2}\epsilon -\partial _{z}^{2}){\mathcal{H}}%
_{\;\;j}^{i}(z,z^{\prime };\mathbf{k}_{\perp },\omega ) &=&\delta
_{\;j}^{i}\delta (z-z^{\prime }), \nonumber \\
\end{eqnarray}%
which implies 
\begin{equation}
{\mathcal{H}}_{\;\;i}^{0}(z,z^{\prime };\mathbf{k}_{\perp },\omega )=0=%
{\mathcal{H}}_{\;\;0}^{i}(z,z^{\prime };\mathbf{k}_{\perp },\omega )
\end{equation}%
and leaves us with only 
\begin{eqnarray}
\hspace{-0.5cm} \epsilon \left( \mathbf{k}_{\perp }^{2}-\omega ^{2}\epsilon -\partial
_{z}^{2}\right) {\mathcal{H}}_{\;\;0}^{0}(z,z^{\prime };\mathbf{k}_{\perp
},\omega ) &=&\delta (z-z^{\prime }), \\
\hspace{-0.5cm} (\mathbf{k}_{\perp }^{2}-\omega ^{2}\epsilon -\partial _{z}^{2}) {\mathcal{H}}%
_{\;\;j}^{i}(z,z^{\prime };\mathbf{k}_{\perp },\omega ) &=&\delta
_{\;j}^{i}\delta (z-z^{\prime }).
\end{eqnarray}%
This system can be solved in terms of the function ${\mathcal{F}_0}(z,z^{\prime };\mathbf{k_{\perp }},\omega )$ satisfying 
\begin{equation}
\left( \mathbf{k}_{\perp }^{2}-\omega ^{2}\epsilon -\partial _{z}^{2}\right)
 {\mathcal{F}_0}(z,z^{\prime };\mathbf{k_{\perp }},\omega )=\delta (z-z^{\prime }),
\end{equation}%
plus the standard BCs at infinity. The result is \cite{Schwinger}
\begin{equation}
{\mathcal{F}_0}(z,z^{\prime };\mathbf{k_{\perp }},\omega )=\frac{ie^{ik_{z}|z-z^{%
\prime }|}}{2k_{z}},
\label{GFRED00}
\end{equation}
with 
\begin{equation}
k_{z}=\left\{ 
\begin{array}{rl}
\sqrt{\tilde{k}_{0}^{2}-\mathbf{k}_{\perp }^{2}} & ,\;\text{if}\;\tilde{k}%
_{0}>\|{\mathbf k}_{\perp }\|, \\ 
i\sqrt{\mathbf{k}_{\perp }^{2}-\tilde{k}_{0}^{2}} & ,\;\text{if}\;\tilde{k}%
_{0}<\|{\mathbf k}_{\perp }\|\;,%
\end{array}%
\right.  \label{kz}
\end{equation}%
where we define $\tilde{k}_{0}=\omega \sqrt{\epsilon }$ with $\omega > 0$
and  $\tilde{k}^{\alpha }=(\sqrt{\epsilon }\omega ,\mathbf{k}_{\perp
},0),\;\tilde{k}^{\alpha }\tilde{k}_{\alpha }=\epsilon \omega ^{2}-\mathbf{k}%
_{\perp }^{2}=\tilde{k}^{2}$. In this way, we obtain
\begin{eqnarray}
&& {\mathcal{H}}_{\;\;0}^{0}(z,z^{\prime };\mathbf{k}_{\perp },\omega )=\frac{1}{%
\epsilon } {\mathcal{F}_0}(z,z^{\prime };\mathbf{k_{\perp }},\omega ),\\
&& {\mathcal{H}}%
_{\;\;j}^{i}(z,z^{\prime };\mathbf{k}_{\perp },\omega )= {\mathcal{F}_0}(z,z^{\prime };%
\mathbf{k_{\perp }},\omega )\delta _{\;j}^{i}.
\end{eqnarray}%
Then, Eq. (\ref{Eq GF temporal Red 1}) can be integrated using $
 {\mathcal{H}}_{\;\;\sigma }^{\nu }(z,z^{\prime })$ as follows 
\begin{eqnarray}
&&  g_{\;\;\sigma }^{\nu }(z,z^{\prime };\mathbf{k}_{\perp },\omega )=
{\mathcal{H}}_{\;\;\sigma }^{\nu }(z,z^{\prime })\nonumber \\
 && -\int dz^{\prime \prime }%
 {\mathcal{H}}_{\;\;\beta }^{\nu }(z,z^{\prime \prime })i\tilde{\theta}\delta
(z^{\prime \prime })\varepsilon _{\quad \;\;\;\gamma }^{3\beta \alpha
}k_{\alpha }g_{\;\;\sigma }^{\gamma }(z^{\prime \prime },z^{\prime };\mathbf{k}%
_{\perp },\omega ), \nonumber \\
\end{eqnarray}
yielding 
\begin{eqnarray}
&&g_{\;\;\sigma }^{\nu }(z,z^{\prime };\mathbf{k}_{\perp },\omega ) =
 {\mathcal{H}}_{\;\;\sigma }^{\nu }(z,z^{\prime })- \nonumber \\
&& \hspace{1.0cm} {\mathcal{H}}_{\;\;\beta
}^{\nu }(z,0)i\tilde{\theta}\varepsilon _{\quad \;\;\;\gamma }^{3\beta \alpha
}k_{\alpha }\;g_{\;\;\sigma }^{\gamma }(0,z^{\prime };\mathbf{k}_{\perp
},\omega ).  \label{RED_GF} 
\end{eqnarray}%
An example of the flexibility of this method is given in Ref. \cite{UrrutiaMartinCambiaso3}, where the free GF was chosen to
satisfy the required BCs in order to limit the space in direction $z$ to the
region between two metallic plates parallel to the interface. Equation (\ref{RED_GF}) reduces the calculation of $g_{\;\;\sigma }^{\nu }$ to solving a set of linearly coupled algebraic equations. The details of the procedure for solving Eq.(\ref{RED_GF}) are presented in
Ref. \cite{UrrutiaMartinCambiaso}. 
The solution is 
\begin{eqnarray}
g_{\;0}^{0}\left( z,z^{\prime };\mathbf{k}_{\perp },\omega \right) &=&\frac{1}{\epsilon }\bar{g}_{\;\;0
}^{0}(z,z^{\prime };\mathbf{k}_{\perp },\omega )\;,\nonumber\\
g_{\;\; \mu}^{i}\left( z,z^{\prime };\mathbf{k}_{\perp },\omega \right)&=&
\bar{g}_{\;\; \mu}^{i}(z,z^{\prime };\mathbf{k}_{\perp },\omega )\;,\\
g_{\;\; i}^{\mu}\left( z,z^{\prime };\mathbf{k}_{\perp },\omega \right) &=&
\bar{g}_{\;\;i}^{\mu}(z,z^{\prime };\mathbf{k}_{\perp },\omega ),\nonumber
\label{RELg_gBAR}
\end{eqnarray}%
where%
\begin{eqnarray}
\bar{g}_{\;\;\nu }^{\mu }(z,z^{\prime };\mathbf{k_{\perp }},\omega ) &=&\eta
_{\;\;\nu }^{\mu } {\mathcal{F}_0}(z,z^{\prime };\mathbf{k_{\perp }},\omega)\nonumber\\
&&+i\varepsilon _{\;\;\nu }^{\mu \;\;\alpha 3}k_{\alpha }P(z,z^{\prime };%
\mathbf{k_{\perp }},\omega )  \notag \\
&&+\tilde{\theta} {\mathcal{F}_0}(0,0;\mathbf{k_{\perp }},\omega )P(z,z^{\prime };\mathbf{k_{\perp }},\omega )\nonumber\\
&&\times\left[ k^{\mu }k_{\nu}-\left( \eta _{\;\;\nu }^{\mu }+u^{\mu }u_{\nu }\right) k^{2}\right],\label{GF temporal Red}
\end{eqnarray}%
and 
\begin{equation}
P(z,z^{\prime };\mathbf{k_{\perp }},\omega )=-\tilde{\theta}\,\frac{%
{\mathcal{F}_0}(z,0;\mathbf{k_{\perp }},\omega ){\mathcal{F}_0}(0,z^{\prime };\mathbf{k_{\perp }}%
,\omega )}{\epsilon -\tilde{\theta}^{2}\tilde{k}^{2}{\mathcal{F}_0}^{2}(0,0;\mathbf{%
k_{\perp }},\omega )},  \label{PREDGF}
\end{equation}%
with $u^{\mu }=(0,0,0,1)$. It is interesting to observe that the term  $P(z,z^{\prime };\mathbf{k_{\perp }},\omega )$ is proportional to  $e^{ik_z(|z|+|z'|)}$, whose  phase is  finally   responsible for the reversed VC radiation, as we will show in Sec. \ref{RCRSEC}. 

Since the only difference between the   GFs 
$G^\mu{}_\nu$ and ${\bar G}^\mu{}_\nu$ is that $G^0{}_0={\bar G}
^0{}_0/\epsilon $, with all other terms being equal, it is convenient to present the barred GFs that are obtained once $\bar{g}_{\;\;\nu }^{\mu }(z,z^{\prime };\mathbf{k_{\perp }},\omega )$ is substituted in Eq. (\ref{GF temporal coord}). {The
full GF matrix $\bar{G}_{\;\;\nu }^{\mu }(\mathbf{x},\mathbf{x}^{\prime };\omega )$ can finally be written as the sum of three
terms}, $\bar{G}_{\;\;\nu }^{\mu }(\mathbf{x},\mathbf{x}^{\prime };\omega )=\bar{G}_{ED\;\nu
}^{\mu }(\mathbf{x},\mathbf{x}^{\prime };\omega )+\bar{G}_{\tilde{\theta}\;\nu }^{\mu
}(\mathbf{x},\mathbf{x}^{\prime };\omega )+\bar{G}_{\tilde{\theta}^{2}\;\nu }^{\mu
}(\mathbf{x},\mathbf{x}^{\prime };\omega )$, where the three pieces are
\begin{eqnarray}
\hspace{-.4cm}\bar{G}_{ED\;\nu }^{\mu }(\mathbf{x},\mathbf{x}^{\prime };\omega )&=&\eta
_{\;\;\nu }^{\mu }4\pi \int \frac{d^{2}\mathbf{k_{\perp }}}{(2\pi )^{2}}e^{i%
\mathbf{k_{\perp }}\cdot \mathbf{R}_{\perp }}\frac{ie^{i\sqrt{\tilde{k}%
_{0}^{2}-\mathbf{k}_{\perp }^{2}}|z-z^{\prime }|}}{2\sqrt{\tilde{k}_{0}^{2}-%
\mathbf{k}_{\perp }^{2}}},  \nonumber \\
\hspace{-.4cm} \bar{G}_{\tilde{\theta}\;\nu }^{\mu }(\mathbf{x},\mathbf{x}^{\prime };\omega
) &=&i\varepsilon _{\;\;\nu }^{\mu \;\;\alpha 3}\frac{4\pi \tilde{\theta}}{4n^2+%
\tilde{\theta}^{2}} \int \frac{d^{2}\mathbf{k_{\perp }}}{(2\pi )^{2}}
e^{i\mathbf{k_{\perp }}\cdot \mathbf{R}_{\perp }}k_{\alpha }\nonumber\\
&&\times\frac{e^{i\sqrt{\tilde{k}_{0}^{2}-\mathbf{k}_{\perp }^{2}}(|z|+|z^{\prime }|)}}{\tilde{k}%
_{0}^{2}-\mathbf{k}_{\perp }^{2}}\;, \nonumber \\
\hspace{-0.4cm}\bar{G}_{\tilde{\theta}^{2}\;\nu }^{\mu }(\mathbf{x},\mathbf{x}^{\prime
};\omega)&=&\frac{ 4\pi i \, {\tilde \theta}^{2}}{4n^2+\tilde{\theta}^{2}} \int \frac{%
d^{2}\mathbf{k_{\perp }}}{(2\pi )^{2}} \nonumber \\
&&\times \left[ k^{\mu }k_{\nu }-\left( \eta
_{\;\;\nu }^{\mu }+u^{\mu }u_{\nu }\right) k^{2}\right] \nonumber \\ && \times  e^{i\mathbf{k_{\perp}}\cdot \mathbf{R}_{\perp }}
\frac{e^{i\sqrt{\tilde{k}_{0}^{2}-\mathbf{k}_{\perp }^{2}}(|z|+|z^{\prime }|)}}
{2\left( \tilde{k}_{0}^{2}-\mathbf{k}_{\perp }^{2}\right) ^{3/2}}.\hspace{1cm}   \label{GF oscillating}
\end{eqnarray}
Let us  emphasize that the potentials and the resulting electromagnetic fields must be calculated using the GF ${G}_{\;\;\nu }^{\mu }(\mathbf{x},\mathbf{x}^{\prime };\omega )$. In the static limit ($\omega=0$), the result (\ref{GF oscillating})
reduces to that reported in Refs. \cite{UrrutiaMartinCambiaso}.

The next step is to evaluate the GFs (\ref{GF oscillating}) in the far-field approximation corresponding to   the coordinate conditions
\begin{eqnarray}
\|\mathbf{x}\|\gg\|\mathbf{x}%
^{\prime}\|, && \,\,\,   R_{\perp}=\|\left(\mathbf{x}-\mathbf{x}%
^{\prime}\right)_{\perp}\|\simeq\|\mathbf{x}_{\perp}\|=\rho, \nonumber \\
&& |z|+|z^{\prime}|\simeq|z|, 
\label{FFC0}
\end{eqnarray}
where $\|\mathbf{x}\|=r \rightarrow \infty $, $\rho \rightarrow \infty$, and $z \rightarrow \infty$. 
In this approximation,  the integrals in Eqs. (\ref{GF oscillating}) include rapidly oscillating exponential functions  whose leading contributions are calculated in the stationary phase approximation and subsequently verified by the steepest descent method  \cite{Chew,Chew1,Mandel}. Also we make repeated use of  
the generic Sommerfeld identity \cite{Sommerfeld}
\begin{eqnarray}
i\int_{0}^{\infty} 
\frac{ k_{\perp} dk_{\perp}}
{\sqrt{\tilde{k}_{0}^{2}-k_{\perp}^{2}}} J_{0}( k_{\perp}R_{\perp})
e^{i\sqrt{\tilde{k}_{0}^{2}-k_{\perp}^{2}} |Z|}=
\frac{e^{i \tilde{k}_{0}{\cal R}}}{\cal R}
\label{SOMM_ID0}
\end{eqnarray}
where  ${\cal R}=\sqrt{R_{\perp}^{2}+Z^{2}}$. We consider mainly two cases: (i)  $Z=|z-z'|$  where  ${\cal R}$ is denoted by $R$ and (ii)
$Z=|z|+|z'|$ where  ${\cal R}$ is denoted by ${\tilde R}$.
The detailed calculation is presented in the Appendix  and the results in the far-field approximation are 
\begin{eqnarray}
\bar{G}_{ED\;\nu }^{\mu }(\mathbf{x},\mathbf{x}^{\prime };\omega ) &=& \eta
_{\;\;\nu }^{\mu }\frac{e^{i\tilde{k}_{0}r}}{r}e^{-i\tilde{k}_{0}\hat{%
\mathbf{n}}\cdot \mathbf{x}^{\prime }}, \label{GF0} \\
 \bar{G}_{\tilde{\theta}\;\nu }^{\mu }(\mathbf{x},\mathbf{x}^{\prime };\omega
) &=&  \varepsilon _{\;\;\nu }^{\mu \;\;\alpha 3}\frac{2\tilde{\theta}}{%
4n^{2} +\tilde{\theta}^{2}}\frac{s_{\alpha }}{|\cos \theta |}\frac{%
e^{i\tilde{k}_{0}r}}{r} \label{GFT}\nonumber\\
&&\times e^{i\tilde{k}_{0}\left( -\mathbf{n}_{\perp }\cdot 
\mathbf{x}_{\perp }^{\prime }+|z^{\prime }\cos \theta |\right) },\\
 \bar{G}_{\tilde{\theta}^{2}\;\nu }^{\mu }(\mathbf{x},\mathbf{x}^{\prime
};\omega )&=& \frac{\tilde{\theta}^{2}}{4n^{2} +\tilde{\theta}^{2}}\frac{%
e^{i\tilde{k}_{0}r}}{r\cos^{2}\theta}C^{\mu }_{\;\;\nu }(\mathbf{x},n)\nonumber\\
&&\times e^{i\tilde{k}_{0}\left( -\mathbf{n}_{\perp }\cdot 
\mathbf{x}_{\perp }^{\prime }+|z^{\prime }\cos \theta |\right)},
\label{GFT2} 
\end{eqnarray}
where we define
\begin{equation}
C^{\mu }_{\;\;\nu }(\mathbf{x},n)=\left( 
\begin{array}{cccc}
\sin ^{2}\theta \quad & -{x}/{rn}&-{y}/{rn}  & 0 \\ 
{x}/{rn}& -{1}/{n^{2}} & 0 & 0 \\ 
{y}/{rn}& 0 & -{1}/{n^{2}}+\sin^{2}\theta \quad  & 0 \\ 
0 & 0 & 0 & 0%
\end{array}%
\right).
\label{MATRIX}
\end{equation}
Here ${s}_\alpha =(1/n, \hat{\mathbf{n}} )$, 
$\hat{\mathbf{n}}$ is a unit vector in the direction of $\mathbf{x}$, and  $%
\mathbf{n}_{\perp }=(x/r, \, y/r \, ,0)=\sin\theta (\cos\phi, \, \sin \phi,\, 0)$ is a vector in the direction of $\mathbf{x}_{\perp
}$ with $\Vert \mathbf{n}_{\perp }\Vert =\sin \theta $. In Eq. (\ref{MATRIX}) we have introduced the shorthand notation $x=r \sin\theta \cos \phi$ and $y=r \sin\theta \sin \phi$ in spherical coordinates.

As explained in the Appendix, the results for the far-field approximation of the  GFs in Eqs. (\ref{GF0})--(\ref{GFT2}) are valid whenever $0< \theta < {\theta}_{\rm C}= \cos^{-1}(1/vn)< \cos^{-1}(1/n) \ll \pi/2$, in such a way that there are no divergences arising from the factors proportional to $1/\cos \theta$ in the GFs, since  $\theta$ is never equal to $\pi/2$.

The GFs (\ref{GF0})--(\ref{GFT2}) together with Eqs. (\ref{DEF_POT}) and  (\ref{A and GF})  yield electromagnetic fields whose Cartesian components  behave like  $(e^{i {\tilde k}_0 r}/r) F(\theta)$ in the far-field approximation. Recalling that ${\tilde k}_0=\omega \sqrt{\epsilon}$ and using  Maxwell equations to leading order in $1/r$  we verify the  expressions 
\begin{equation}
{\mathbf {\hat n}} \cdot {\mathbf E}=0,\quad {\mathbf {\hat n}} \cdot {\mathbf B}=0, \quad  {\mathbf B}=\sqrt{\epsilon} \,  {\mathbf {\hat n}}\times {\mathbf E}, \quad
\label{RHT} 
\end{equation}
which are the distinctive feature of the radiation fields. The three vectors ${\mathbf E}$, ${\mathbf B}$, and ${\mathbf {\hat n}}$  
define a right-handed triad resulting in the  Poynting vector for  a material media with $\mu=1$,
\begin{equation}
{\mathbf S}=\frac{1}{4\pi} {\mathbf E}\times {\mathbf H}=\frac{\sqrt{\epsilon}}{4\pi}|{\mathbf E}|^2 \, {\mathbf {\hat n}},
\label{POYNTING}
\end{equation}
where ${\mathbf {\hat n}}$ is in  the direction of the phase velocity  of the outgoing wave, as appropriate for right-handed materials.

Finally, it is pertinent to emphasize an important difference in the phase of the exponentials related to the source variables ${\mathbf x}'$ in the GFs (\ref{GF0})--(\ref{GFT2}).  In the first case, we encounter the exponential  $e^{i{\tilde k}_0 R}$,  which in the coordinate approximation of the far-field zone produces the phase $i{\tilde k}_0(r-{{\mathbf n}}_\perp\cdot {\mathbf x}'_\perp - z' \cos \theta)$ characteristic of radiation in   standard electrodynamics \cite{Schwinger}. On the other hand, the contributions to the GF proportional to $\tilde \theta$ and $\tilde \theta^2$ involve the exponential $e^{i{\tilde k}_0 {\tilde R}}$  with ${\tilde R}=\sqrt{({\mathbf x}-{\mathbf x}')_\perp^2+(|z|+|z'|)^2}$, which can be presented as follows:
\begin{eqnarray}
{\tilde R}&=&\sqrt{({\mathbf x}-{\mathbf x}')^2-(z-z')^2+(|z|+|z'|)^2}, \nonumber \\
&=& \sqrt{({\mathbf x}-{\mathbf x}')^2 +2(|z z'|+z z')},\nonumber \\
&=& r-{\mathbf n}_\perp\cdot {\mathbf x}'_\perp + |z' \cos \theta|,
\label{RTILDE2}
\end{eqnarray} 
in the far-field approximation. From the second line in Eq. (\ref{RTILDE2}), we remark that whenever the sign of $z z'$ is positive we will have an additional relative phase contributing to the GFs, which will show up in observable quantities as the radiated power, for example. The  term $(|z|+|z'|)^2$ can ultimately be traced back to the form of the reduced GF (\ref{GF temporal Red}) together with expressions  (\ref{GFRED00}) and (\ref{PREDGF}). As we will show in the next section, reversed VC radiation arises precisely due to the contribution $|z' \cos\theta|$ in the phase of the GF   deriving   from the term $(|z|+|z'|)^2$.   
 
\section{The reversed {Vavilov-\v{C}erenkov} radiation }

\label{RCRSEC}

Let us now consider a particle with charge $q$ moving with  constant velocity $v {\mathbf {\hat u}}$,  perpendicular to the interface  $\Sigma$ defined by the $x-y$ plane ($z=0$), as shown in Fig.\ref{REGIONS}. The charge and current densities are
\begin{equation}
 \varrho (\mathbf{x}^{\prime };\omega )=\frac{q}{v}\delta (x^{\prime })\delta(y^{\prime })e^{i\omega \frac{z^{\prime }}{v}},\, \mathbf{J}(\mathbf{x}
^{\prime };\omega )=q\delta (x^{\prime })\delta (y^{\prime })e^{i\omega\frac{z^{\prime }}{v}}\mathbf{\hat{u}},
\end{equation}
 where we henceforth assume $v > 1/\sqrt{\epsilon}>0$.
Instead of an infinite path for the charge, we will take its movement in the interval $z\in
(-\zeta ,\zeta )$, with $\zeta\gg v/\omega$. In the far-field approximation, the resulting components of the  electric field $\mathbf{E}(\mathbf{x}%
;\omega )=-i{\tilde k}_0\mathbf{\hat{n}}A^{0}(\mathbf{x};\omega )+i\omega\mathbf{A}(%
\mathbf{x};\omega )$, calculated in terms of the potential $A^{\mu }(\mathbf{x};\omega )$
via Eq. (\ref{A and GF}), are 
\begin{eqnarray}
E^{1}(\mathbf{x};\omega ) &=&-\sin \theta \frac{i\omega q
e^{i\tilde{k_{0}}r}}{vrn}\Bigg[ \cos\phi\mathcal{I}_{1}(\omega,\theta
) +n\mathcal{I}_{2}(\omega,\theta )\Bigg.\nonumber\\
&&\Bigg.\times\frac{2\tilde{\theta}}{4n^{2}+\tilde{\theta}^{2}}\left(\frac{\sin
\phi }{|\cos \theta |}-\tilde{\theta}\frac{\cos \phi }{2n}\right)\Bigg],\\
E^{2}(\mathbf{x};\omega ) &=&-\sin \theta \frac{i\omega q
e^{i\tilde{k}_{0}r}}{vrn}\Bigg[\sin\phi\mathcal{I}_{1}(\omega,\theta)
-n\mathcal{I}_{2}(\omega,\theta)\nonumber \Bigg.\\
&&\Bigg.\times\frac{2\tilde{\theta}}{4n^{2}+\tilde{\theta}^{2}}\left(
\frac{\cos \phi }{|\cos \theta |}+\tilde{\theta}\frac{\sin \phi}{2n}\right)\Bigg], \\
E^{3}(\mathbf{x};\omega ) &=&\sin\theta\frac{ iq \omega
e^{i\tilde{k}_{0}r}}{vrn}\left[\left( vn-\cos \theta\right)
\frac{\mathcal{I}_{1}(\omega,\theta)}{\sin\theta}\right.\nonumber\\
&&-\left.\frac{2\tilde{\theta}}{4n^{2}+\tilde{\theta}^{2}}\tilde{\theta}
\frac{\mathcal{I}_{2}(\omega,\theta )}{2\cot\theta}\right]. \label{E CR TR}
\end{eqnarray}

\begin{figure}[tbp]
\begin{center}
\includegraphics[scale=0.46]{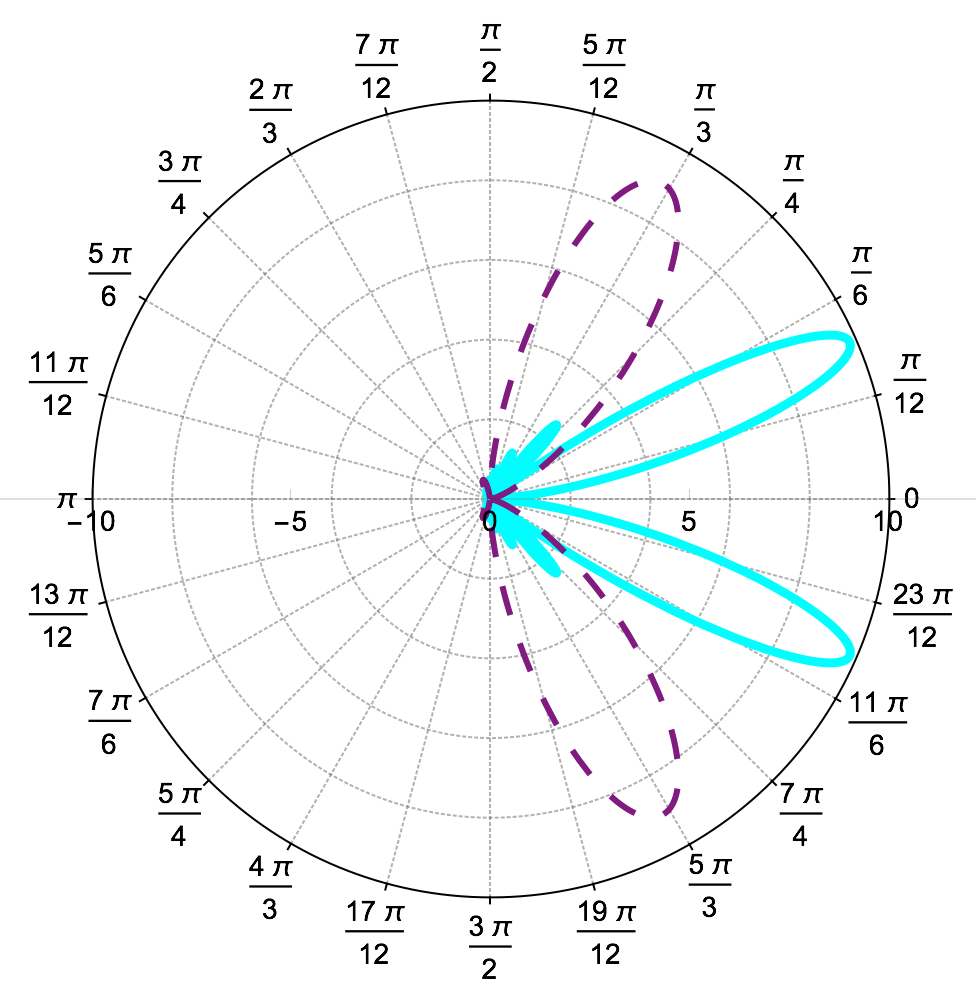}
\end{center}
\caption{{\protect\small Angular distribution  for the  radiated energy per unit frequency  in the case of standard ($\tilde \theta=0$) forward   VC radiation  for $n=2$ and  $\omega=2.48$ eV. The  dashed purple line corresponds to  $v=0.9$ and $\zeta=1.0$ eV$^{-1}$ and the  solid cyan line to  $v=0.5009$ and $\zeta=2.80$ eV$^{-1}$. The scale in the polar axis is in arbitrary dimensions and  runs from zero to ten. The charge moves from left to right. }}
\label{CHE1}
\end{figure}

\begin{figure}[tbp]
\begin{center}
\includegraphics[scale=0.46]{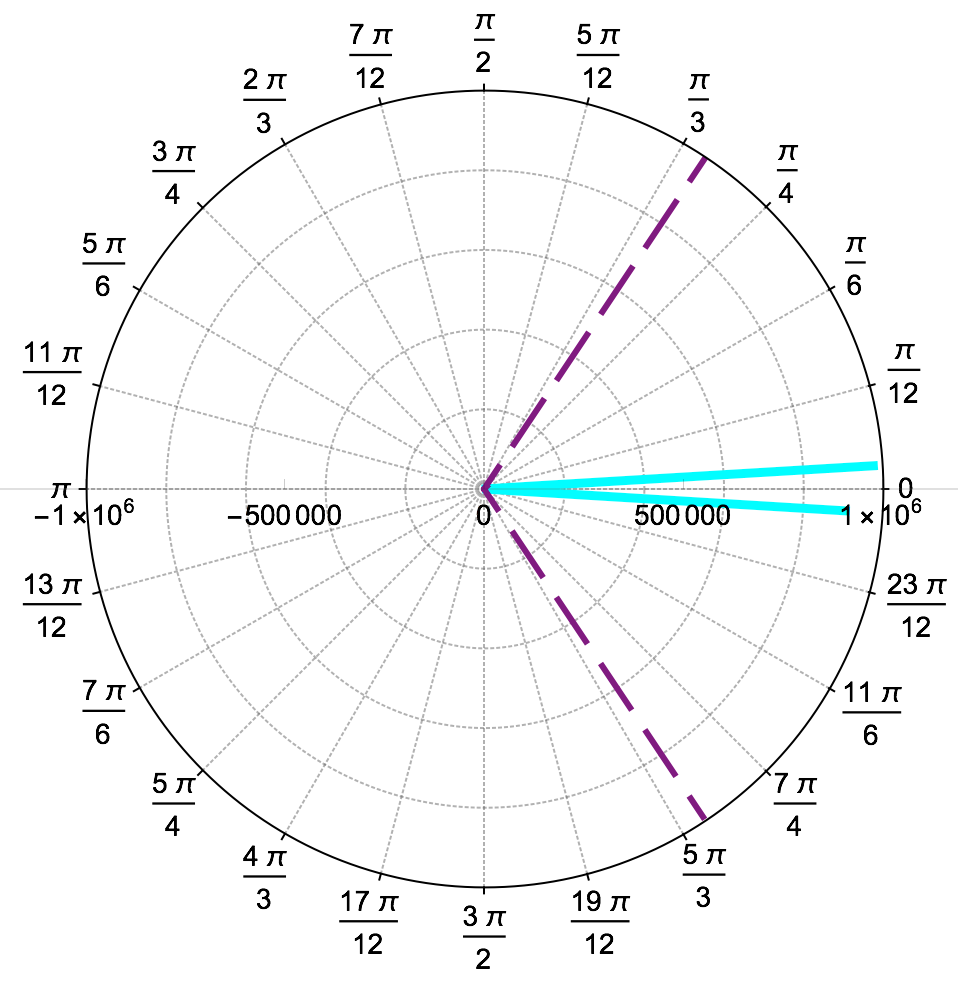}
\end{center}
\caption{{\protect\small Angular distribution for the radiated energy per unit frequency  in the case of standard ($\tilde \theta=0$) VC radiation  for the choices
for $n=2$,  $\omega=2.48$ eV. The  dashed purple line corresponds to  $v=0.9$ and $\zeta=343$ eV$^{-1}$ and the  solid cyan line to  $v=0.5009$ and $\zeta=4830$ eV$^{-1}$.
The scale in the polar axis is in arbitrary dimensions and runs from $0$ to $10^6$. The charge moves from left to right.  }}
\label{CHE12}
\end{figure}

We can verify that $\mathbf{ {\hat n}\cdot E}=0$ as required.  The magnetic field can be obtained from Eq. (\ref{RHT}).
The  integrals $\mathcal{I}_{1}(\omega,\theta )$ and $\mathcal{I}_{2}(\omega,\theta )$  appear when taking the convolution  of the charge and current densities with the GF components  in Eqs. 
(\ref{GF0})--(\ref{GFT2}). They are defined by 
\begin{eqnarray}
&&\hspace{-1cm} \mathcal{I}_{1}(\omega,\theta )=\int_{-\zeta }^{\zeta }dz^{\prime }e^{i%
\frac{\omega z^{\prime }}{v}(1-vn\cos \theta )}=\frac{2\sin\left(\zeta\Xi_{-}\right) }{\Xi_{-} },  \label{I1 function} \\
&&\hspace{-1cm} \mathcal{I}_{2}(\omega,\theta )=\int_{-\zeta }^{\zeta }dz^{\prime
}e^{i{\tilde k}_{0}|z^{\prime }\cos \theta |+i\omega\frac{z^{\prime }}{v}} = \frac{\sin\left(\zeta\tilde{\Xi}_{-}\right) }{\tilde{\Xi}_{-}}\nonumber \\ 
&& \hspace{-0.5cm} +\frac{\sin \left(\zeta\tilde{\Xi}_{+}\right) }{\tilde{\Xi}_{+}} +2i \frac{\sin ^{2}\left( \frac{\zeta }{2}\tilde{\Xi}_{-} \right) }{\tilde{\Xi}_{-} }-2i\frac{%
\sin ^{2}\left(\frac{\zeta}{2}\tilde{\Xi}_{+}\right) }{\tilde{\Xi}_{+}},
\label{I22 function}
\end{eqnarray}%
where 
\begin{eqnarray}
\Xi_{\pm}= \frac{%
\omega }{v}\left( 1\pm vn\cos \theta \right), \quad 
\tilde{\Xi}_{\pm}= \frac{%
\omega }{v}\left( 1\pm vn|\cos \theta | \right). \label{XI}
\end{eqnarray}
In the following, we will learn  that the production of  VC radiation depends on the zeros of $\Xi_{\pm}$ and ${\tilde \Xi}_{\pm}$,  which determine the angle $\theta_C$ of the VC cone. Then it is reasonable to expect that the $|\cos \theta|$ dependence of  ${\tilde \Xi}_{\pm}$ yields  new possibilities. In calculating the right-hand side of Eqs.  (\ref{I1 function}) and (\ref{I22 function}) in the limit $\zeta \gg v/\omega$, which effectively means $\zeta \rightarrow \infty$, we encounter expressions like $\sin(\zeta aN)/(a N)$, which  behave as $\pi \delta(a N)$ \cite{Lighthill}. 
We take advantage of this $\delta$-like behavior
by setting equal to zero all the rapidly oscillating  contributions arising from  those functions with an argument  that can never be zero, like
$\tilde{\Xi}_{+}$, for example. This is relevant to obtain the final expression for  $\mathcal{I}_{2}(\omega,\theta )$, which then simplifies  to  
 \begin{eqnarray}
\mathcal{I}_{2}(\omega,\theta )&=&\frac{\sin\left( \zeta \tilde{\Xi}_{-} \right) }{\tilde{\Xi}_{-} }
+2i\frac{\sin ^{2}\left( \frac{\zeta }{2}\tilde{\Xi}_{-} \right) }{\tilde{\Xi}_{-}}.
\label{I2 function}
\end{eqnarray}%
Recalling  that $\mathbf{B}=\hat{\mathbf{n}}\times \left( n\mathbf{E}%
\right)$ for radiation fields, with $n=\sqrt{\epsilon}$, the angular distribution of the total radiated energy per unit frequency in the interval $-\zeta < z < +\zeta$ with $ \zeta \rightarrow \infty $ 
 is \cite{Schwinger,
Jackson} 
\begin{equation}
\frac{d^{2}E}{d\omega d\Omega }=\frac{nr^{2}}{4\pi ^{2}}\mathbf{E}^{\ast }(%
\mathbf{x};\omega )\cdot \mathbf{E}(\mathbf{x};\omega ),
\end{equation}%
which can be written  as the sum of the following three terms:
\begin{eqnarray}
\frac{d^{2}E_{1}}{d\omega d\Omega } &=&\frac{n\omega^{2}q^{2}}{4\pi ^{2}}\left(
1-\frac{1}{v^{2}n^{2}}\right) \mathcal{I}_{1}^{2}(\omega,\theta ),
\label{d2E 1} \\
\frac{d^{2}E_{12}}{d\omega d\Omega } &=&-\frac{n\omega^{2}q^{2}}{2\pi ^{2}}%
\frac{\tilde{\theta}^{2}}{4n^{2}+\tilde{\theta}^{2}}\left( 1-\frac{1}{%
v^{2}n^{2}}\right) \mathcal{I}_{1}(\omega,\theta )\nonumber\\
&&\times\rm{Re}\left[ \mathcal{I}%
_{2}(\omega,\theta )\right],  \label{d2E 12}\\
\frac{d^{2}E_{2}}{d \omega d\Omega } &=&\frac{n\omega^{2}q^{2}}{4\pi ^{2}}\frac{%
\tilde{\theta}^{2}}{4n^{2}+\tilde{\theta}^{2}}\left( 1-\frac{1}{v^{2}n^{2}}%
\right) \big|\mathcal{I}_{2}(\omega,\theta )\big|^{2}.\nonumber\\ \label{d2E 2}
\end{eqnarray} 
To obtain the above expressions  we have used again the $\delta$-like behavior of the functions ${\cal I}_1$ and ${\cal I}_2$ by  replacing $\theta$ by $\theta_C= \cos^{-1}(1/vn)$ in all the functions of $\theta$ that multiply them.  We have verified the cancellation of  the terms proportional to ${\tilde \theta}^3$. Also, the addition of the contributions  in the term containing the factor $1/(4n^2 +{\tilde \theta}^2)^2$  is such  that the final result ends up being proportional to ${\tilde \theta}^2/(4n^2 + {\tilde \theta^2 )}$. 

The quantity $d^2 E/d\omega d\Omega$ is also referred to as the spectral distribution of the radiation \cite{Schwinger}, a shorter synonymous that we will  use in the following.
Let us observe that the above distributions have azimuthal symmetry and are  even functions of both the angle $\theta$ [recall Eqs. (\ref{I1 function})  and (\ref{I2 function})]  and the MEP   $\tilde{\theta}$. In other words, the leading corrections arising from the ME effect  are of order ${\tilde \theta}^2$. Setting $\tilde{\theta}=0$ we recover the well-known expression for the spectral distribution of the radiation in the standard  VC case \cite{Panofsky}
\begin{equation}
\frac{d^{2}E_{1}}{d \omega d\Omega }=\frac{n\omega^{2}q^{2}}{\pi ^{2}}\left( 1-%
\frac{1}{v^{2}n^{2}}\right) \frac{\sin ^{2}\left(\zeta
\Xi_{-} \right) }{\Xi_{-} ^{2}}.  \label{DEDKDO}
\end{equation}%
Substituting $\mathcal{I}_{1}(\omega,\theta )$ and $\mathcal{I}%
_{2}(\omega,\theta )$ from Eqs. (\ref{I1 function}) and (\ref{I2 function})
 in Eqs. (\ref{d2E 12}) and (\ref{d2E 2}), we have 
\begin{eqnarray}
\frac{d^{2}E_{12}}{d \omega d\Omega } &=&-\frac{n\omega^{2}}{\pi ^{2}}\frac{%
\tilde{\theta}^{2}q^{2}}{4n^{2}+\tilde{\theta}^{2}}\left( 1-\frac{1}{%
v^{2}n^{2}}\right)\nonumber\\
&&\times \frac{\sin \left( \zeta \Xi_{-} \right)\sin\left( \zeta\tilde{\Xi}_{-}\right) }{\Xi_{-}\tilde{\Xi}_{-}}%
,\label{dE 12 explicit}\\
\frac{d^{2}E_{2}}{d \omega d\Omega } &=&\frac{n\omega^{2}}{4\pi ^{2}}\frac{%
\tilde{\theta}^{2}q^{2}}{4n^{2}+\tilde{\theta}^{2}}\left( 1-\frac{1}{%
v^{2}n^{2}}\right)\nonumber\\
&&\times \left[\frac{\sin^{2} \left( \zeta \tilde{\Xi}_{-} \right) }{\tilde{\Xi}_{-}^{2} }+ \frac{\sin ^{4}
\left( \frac{\zeta }{2}\tilde{\Xi}_{-} \right) }{%
\frac{1}{4}\tilde{\Xi}_{-}^{2} }\right].\label{dE 2 explicit}
\end{eqnarray}%
Equations (\ref{DEDKDO})--(\ref{dE 2 explicit}) summarize the spectral distribution of the radiation  when the charges passes through a $\vartheta$ medium. The $\delta$-like behavior of the functions appearing there yields the VC radiation condition   $\sin^2 \theta_{C}= 1-1/( v^2 n^2 )$ and clearly shows that the distributions (\ref{DEDKDO}) and  (\ref{dE 12 explicit}) contribute only to the forward Vavilov-\v{C}erenkov  radiation  with
$\cos \theta_{C}= 1/(nv) > 0$. This is  precisely due to the dependence upon $|\cos \theta|$ in the angular distribution originating from ${\tilde \Xi}_{-} $ in Eq. (\ref{XI}). 

Even though the spectral distribution in Eq. (\ref{dE 12 explicit}) depends on ${\tilde \Xi}_{-}$, its contribution  shows up in  a product that behaves like  $\delta(1-vn \cos \theta)\times \delta(1-vn|\cos\theta|)$ in the limit of large $\zeta$. This yields a nonzero spectral distribution $d^2 E_{12}/d\omega d\Omega $ only when $\cos \theta >0$, i.e., in the forward direction. 
On the contrary, the distribution $d^2 E_2/d\omega d \Omega$  contributes both to the forward VC cone
($\cos \theta >0$), as well as to the backward VC cone ($\cos \theta < 0$) because ${\tilde \Xi}_{-}$, depending upon $|\cos \theta|$, admits also a zero in the range $\pi/2 < \theta < \pi$. 
In other words, radiation is also detected in the backward direction, i.e.,  $\cos \theta < 0 $, according to the angular distribution (\ref{dE 2 explicit}). This  defines the reversed VC cone and   constitutes the most important result of the manuscript. 

The forward VC radiation receives corrections of order ${\tilde \theta}^2$ with respect to the standard case.  The reversed VC radiation is purely of order ${\tilde \theta}^2$ and, in general, will be strongly suppressed with respect to the forward output, however, it   is different from zero. If the angle of the forward  cone is $\theta_0$, the one  corresponding to the reversed cone will be $\pi-\theta_0$. Since this is standard electrodynamics plus  additional field-dependent  sources at the interface, the group and phase velocities of  the electromagnetic wave are parallel, as already pointed out after Eq.(\ref{POYNTING}). In fact this model provides the usual interpretation of the radiation of a system observed at an arbitrary solid angle $d \Omega$ \cite{Schwinger, Jackson}.

Figures \ref{CHE1} and \ref{CHE12} show the spectral distribution  in the case of pure forward VC radiation, Eq. (\ref{DEDKDO}), in a dielectric with $n=2$ and $\tilde \theta=0$, for the average frequency of  $\omega=2.5$ eV ($500$ nm) in the VC radiation  spectrum. In each figure, the particle moves from left to right along the line ($\pi-0$). In Fig. \ref{CHE1}, the dashed purple line corresponds to  $v= 0.9$ and $\zeta=1.0$ eV$^{-1}$ and  the solid cyan line to $v= 0.5009$ and $\zeta= 2.8$ eV$^{-1}$. Such low values of $\zeta$ are chosen to illustrate the emergence of the forward  radiation lobes. In Fig. \ref{CHE12},  the dashed purple line corresponds to  $v= 0.9$ and $\zeta=340$ eV$^{-1}$ and the solid cyan line to  $v= 0.5009$ and $\zeta= 4800$ eV$^{-1}$. Such an increase in each  value of  $\zeta$ is enough to show the appearance of the  forward VC cone in both  cases. 

\begin{figure}[tbp]
\begin{center}
\includegraphics[scale=0.46]{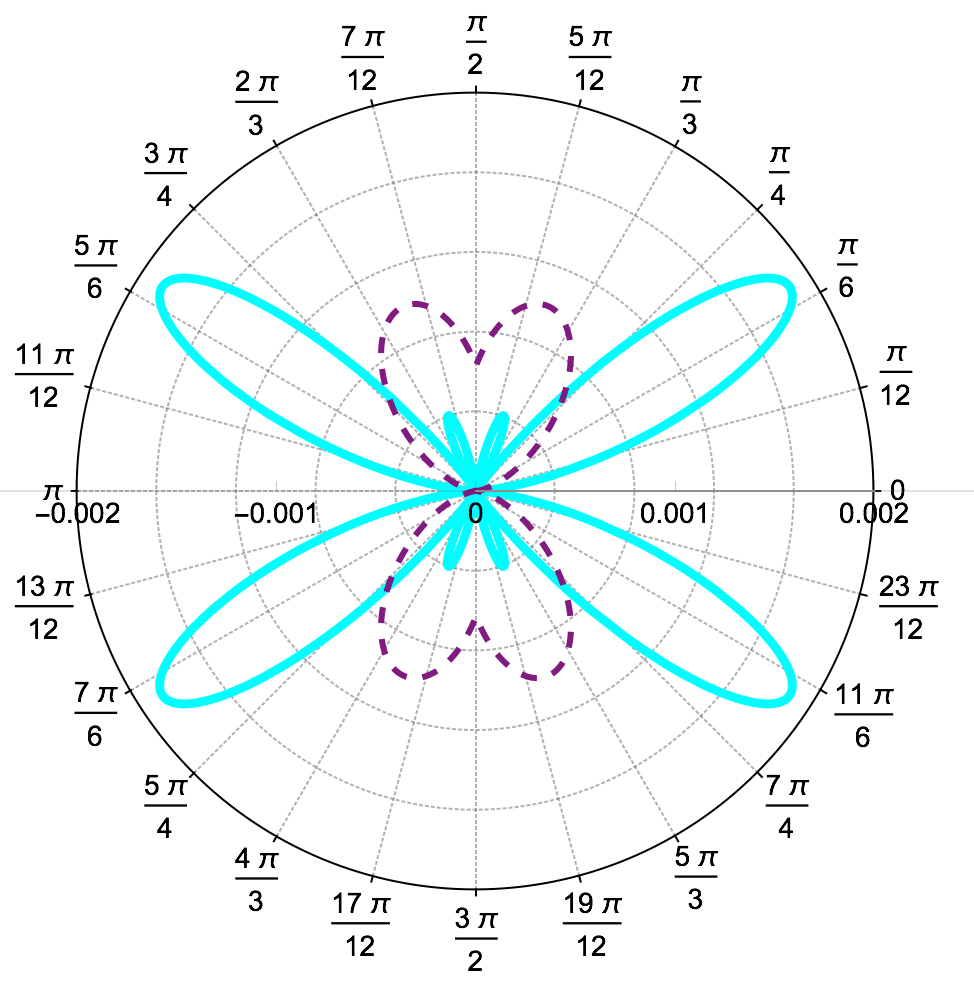}
\end{center}
\caption{{\protect\small Angular distribution  for the  radiated energy per unit frequency  in the case of reversed VC radiation 
for $n=2$, $\omega=2.48$ eV and $\tilde{\theta}=11\alpha$. The
 solid cyan line corresponds to  $v=0.5009$ and $\zeta=2.8$ eV$^{-1}$ and the dashed purple line   to $v=0.9$ and $\zeta=1.0$ eV$^{-1}$.
The scale in the polar axis is in arbitrary dimensions and runs from $0$ to $2\times 10^{-3}$. The charge moves from left to right.}}
\label{CHE2}
\end{figure}

\begin{figure}[tbp]
\begin{center}
\includegraphics[scale=0.46]{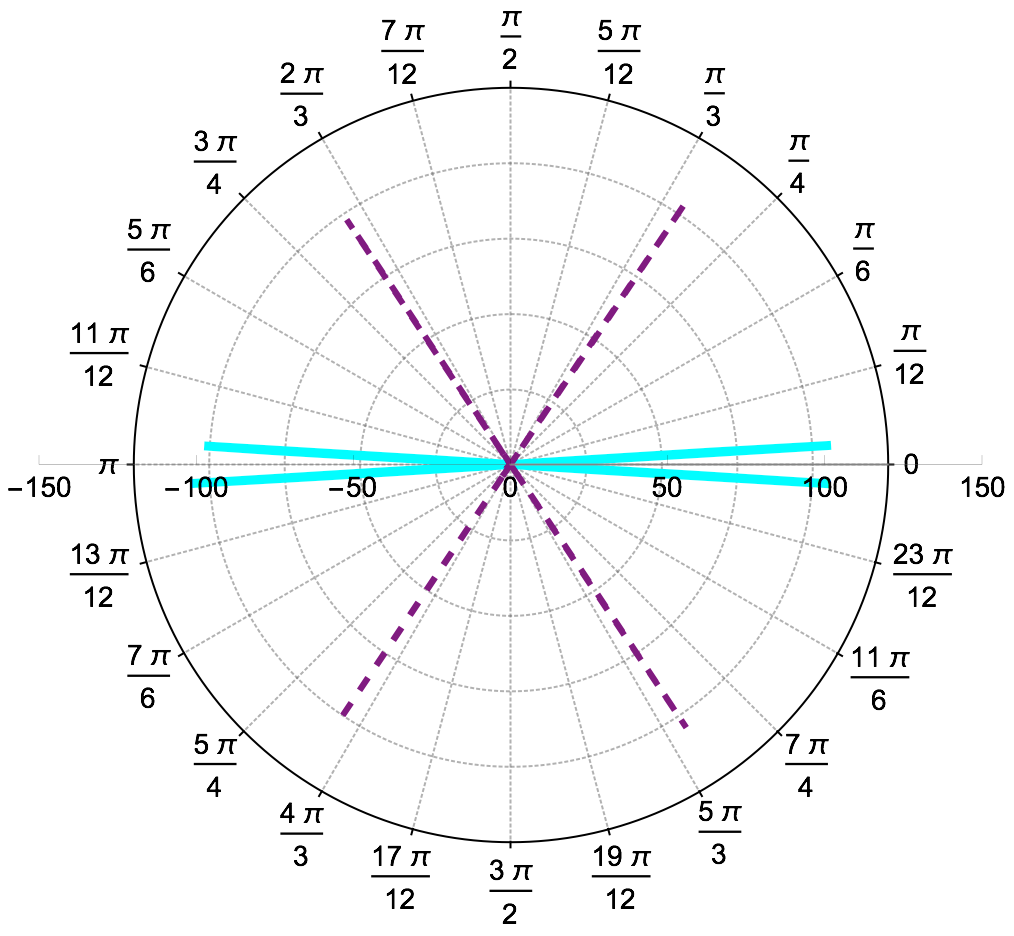}
\end{center}
\caption{{\protect\small Angular distribution  for the radiated energy per unit frequency  in the case of reversed VC radiation 
for $n=2$, $\omega=2.48$ eV and $\tilde{\theta}=11\alpha$. 
The solid cyan line corresponds to  $v=0.5009$ and $\zeta=4830$ eV$^{-1}$, and the dashed purple line   to  $v=0.9$ and $\zeta=343$ eV$^{-1}$. The scale in the polar axis is in arbitrary dimensions and runs from $0$ to $ \sim 10^2$. The charge moves from left to right.}}
\label{CHE21}
\end{figure}

Figures \ref{CHE2} and \ref{CHE21} include the spectral distribution of the radiation  arising from the contribution of Eq. (\ref{dE 2 explicit}) in a medium with ${\tilde \theta} \neq 0$, which clearly show  the presence of  reversed VC radiation . This term  also provides corrections to the forward VC radiation. Both additions are highly suppressed with respect to the forward VC radiation, so that  the scales in both figures are separately   chosen such as to  make these small, but nonzero, contributions clearly visible.
Here we take again the frequency of $\omega=2.5$ eV and  the charge propagates now in a  $\vartheta$ medium with $n=2$ and $\tilde \theta= 11 \alpha$, from left to right  along the line ($\pi-0$). In Fig. \ref{CHE2}, the dashed purple line corresponds again to $v= 0.9$ and $\zeta=1.0$ eV$^{-1}$ and  the solid cyan line to $v= 0.5009$ and $\zeta= 2.8$ eV$^{-1}$.  The contribution of 
$d^2E_2/d\omega d\Omega$ to the  forward and reversed VC cones are displayed in Fig. \ref{CHE21}, where the parameters $\zeta$ have been  modified  with respect  to those
 in  Fig. 4 by increasing them   to $343$ eV$^{-1}$ (dashed purple line) and $4830$ eV$^{-1}$ (solid cyan line), respectively. All the spectral  distributions we have plotted are calculated  from the respective expressions $d^2 E/d \omega d \Omega $ and  the results are expressed  in units of the common factor $q^2/\pi^2= 7.4 \times 10^{-4}$ for $q=\sqrt{\alpha}$.
 Let us emphasize that the negative contribution in Eq. (\ref{dE 12 explicit}) only diminishes the radiation in the forward direction but does not affect the reversed VC radiation. 

\begin{widetext}

\begin{figure}[tbp]
\begin{center}
\includegraphics[scale=0.40]{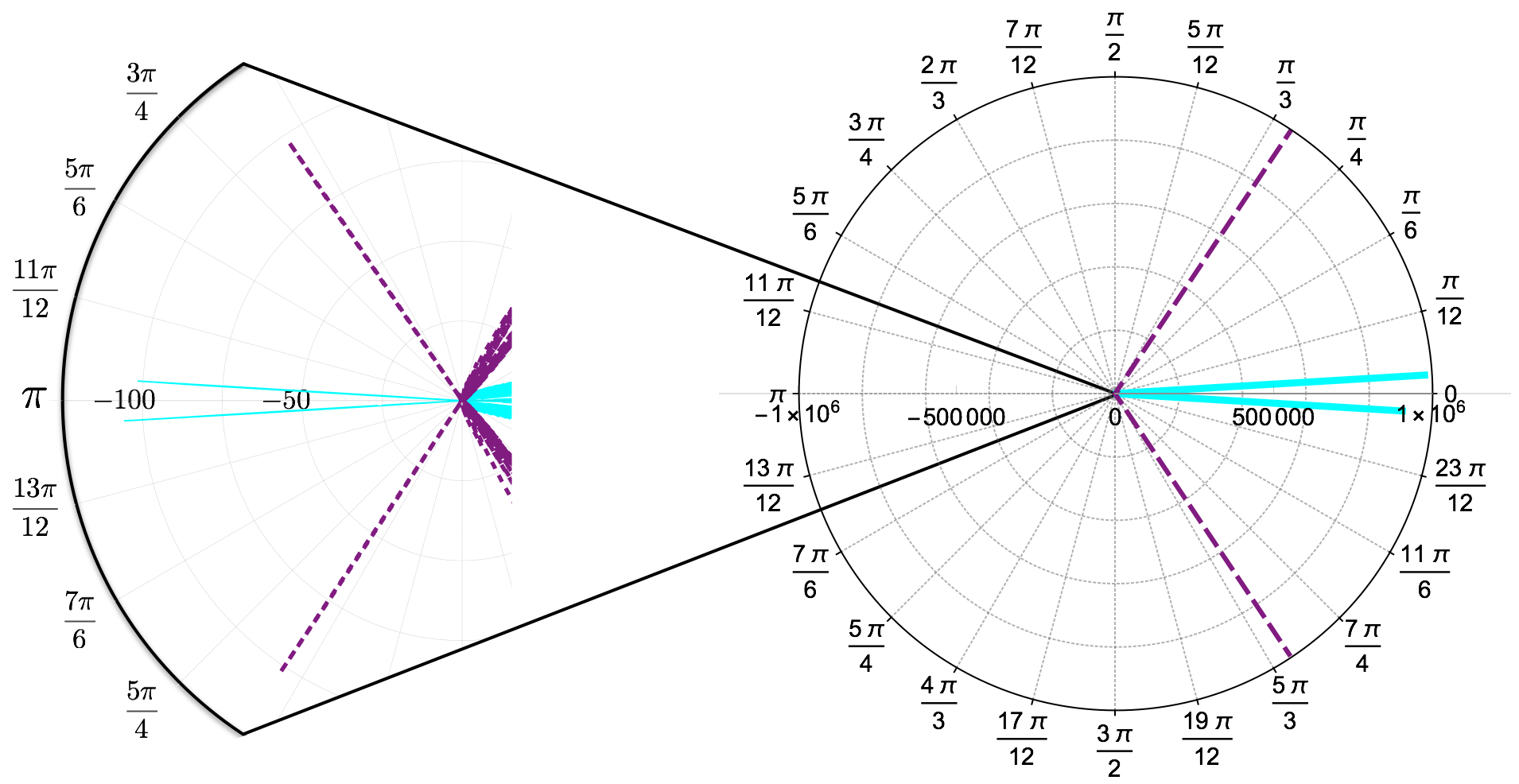}
\end{center}
\caption{{\protect\small Angular distribution for the total radiated energy per unit frequency  for the full VC radiation  for the choices
for $n=2$,  $\omega=2.48$ eV  and ${\tilde \theta}=11 \alpha$. The  dashed purple line corresponds to  $v=0.9$ and $\zeta=343$ eV$^{-1}$ and the  solid cyan line to  $v=0.5009$ and $\zeta=4830$ eV$^{-1}$. The scale of the polar plot is in arbitrary dimensions and runs  from $0$ to $10^6$. On the left side of the figure we plot an elargement in the backward direction showing the onset of the reversed VC radiation. Here the radial scale goes from 0 to $10^2$. The charge moves from left to right. }}
\label{FULLCHE}
\end{figure}

\end{widetext}

The full spectral distribution  of the total radiated energy is given by the sum of Eqs. (\ref{DEDKDO})--(\ref{dE 2 explicit}) and it is plotted  in Fig. \ref{FULLCHE}. The scale of the polar plot is in arbitrary dimensions and runs  from $0$ to $10^6$. On the left side of the figure we plot an elargement in the backward direction showing the onset of the reversed VC radiation. Here the radial scale goes from 0 to $10^2$ showing the high suppression of the  radiation in the backward direction  with respect to that in the forward direction.

\section{Total radiated energy}

\label{Total E}

In this section, we calculate the total energy per unit frequency  radiated  by the charge on its path from $-\zeta$ to $+\zeta$. Let us first review the calculation of ${dE_{1}}/{d\omega}$, following  the procedure in Ref.
\cite{Panofsky}.  Integrating expression (\ref{DEDKDO}) with respect to the solid angle we obtain   
\begin{equation}
\frac{dE_{1}}{d\omega}=\frac{2n \omega^{2}q^{2}}{\pi}\left(1-\frac{1}{v^{2}n^{2}}\right)\int_{-1}^{1}d(\cos\theta)
\frac{\sin^{2}\left(\zeta\Xi_{-}\right)}{\Xi_{-}^{2}}.
\end{equation}
The $\delta$-like behavior  of the integrand in the limit $\zeta \gg \omega/v $ shows that the radiation is sharply localized in a main lobe around the angle $\theta_C $  given by $\cos \theta_C=1/(vn)$, yielding a contribution only in the forward direction.
Therefore, making the change of variable $u=\frac{ \omega \zeta}{v}\left(1-vn\cos\theta\right)$,   we can safely extend the integration limits of $u$ to $\pm \infty$ as long as we include the maximum  located at $u=0$. The result is
\begin{eqnarray}
\frac{dE_{1}}{d \omega}&=&\frac{2\omega q^{2}\zeta}{\pi}\left(1-\frac{1}{%
v^{2}n^{2}}\right)\int_{-\infty}^{\infty}du\frac{\sin^{2}u}{u^{2}},\nonumber\\
&=& q^{2}\omega L \left(1-\frac{1}{v^{2}n^{2}}\right),
\label{DEDOMEGAST}
\end{eqnarray}
where we have introduced  the total length $L=2\zeta$ traveled by the particle, thus  recovering  the standard result \cite{Panofsky}. In other words, we are estimating the contribution  from   each sharply localized lobe as 
\begin{equation}
\int_{\rm lobe}d(cos \theta)\frac{\sin^{2}\left(\zeta\Xi_{-}\right)}{\Xi_{-}^{2}}=\frac{\zeta}{\omega n}\int_{-\infty}^{+\infty} du \frac{\sin^2u}{u^2}=\frac{\zeta \pi}{\omega n}.
\label{LOBE_EST}
\end{equation}

The next term comes from the angular integration of Eq. (\ref{dE 12 explicit}),
\begin{eqnarray}
\frac{dE_{12}}{d \omega}&=&- \frac{n\omega^{2}}{\pi ^{2}}\frac{\tilde{\theta}%
^{2}q^{2}}{4n^{2}+\tilde{\theta}^{2}}\left( 1-\frac{1}{v^{2}n^{2}}\right)\nonumber\\
&& \times\int d \Omega \,
\frac{\sin \left(\zeta \Xi_{-} \right)\sin ( \zeta\tilde{\Xi}_{-} )
}{\Xi_{-}\tilde{\Xi}_{-} }.
\end{eqnarray}
The presence of $|\cos \theta|$ from the function $(\zeta\tilde{\Xi}_{-})$ in the integrand requires separating the integration with respect to $\theta$ into the regions $0<\theta < \pi/2 \, \,$ and $ \,\, \pi/2 < \theta < \pi$.
After splitting the integral as required  and setting equal to zero  the rapidly oscillating
contribution proportional to $\sin \left(\zeta\Xi_{-} \right)$ in the region $\pi/2 < \theta < \pi$, we are left with   
\begin{eqnarray}
\frac{dE_{12}}{d \omega}&=&- \frac{\tilde{\theta}^{2}q^{2}}{4n^{2}+\tilde{\theta}%
^{2}}\frac{2n \omega^{2}}{\pi }\left( 1-\frac{1}{v^{2}n^{2}}\right) \nonumber \\
&& \times \int_{0}^{1}d(\cos \theta )\frac{\sin ^{2}\left(\zeta\Xi_{-} \right) }{\Xi_{-} ^{2}}.
\end{eqnarray}
Again, this contribution to the  radiation is only in the forward direction and it is concentrated in a main lobe around $\theta_C$.  Making use of Eq. (\ref{LOBE_EST}) yields
\begin{eqnarray}
\frac{dE_{12}}{d \omega}&=&-\frac{\tilde{\theta}^{2}q^{2}}{4n^{2}+\tilde{\theta}%
^{2}}\frac{2 \omega\zeta }{\pi }\left( 1-\frac{1}{v^{2}n^{2}}\right)  
\int_{-\infty}^{\infty }du\frac{\sin ^{2}u}{u^{2}},\nonumber\\
&=&-\frac{\tilde{\theta}^{2}q^{2}}{%
4n^{2}+\tilde{\theta}^{2}}\omega L \left( 1-\frac{1}{v^{2}n^{2}}\right).
\end{eqnarray}

The calculation of  ${dE_{2}}/{d\omega}$ starts from Eq. (\ref{dE 2 explicit}), whose separate contributions are 
\begin{eqnarray}
&&\hspace{-1cm}\frac{d^{2}E_{2_{a}}}{d\omega d\Omega } 
=\frac{n\omega^{2}}{4\pi ^{2}}\frac{\tilde{\theta}^{2}q^{2}}{4n^{2}+\tilde{\theta}^{2}}\left( 1-\frac{1}{v^{2}n^{2}}\right)\frac{\sin^{2} \left( \zeta\tilde{\Xi}_{-} \right) }{\tilde{\Xi}_{-}^{2}},\label{d2E 2 Re explicit}\\
&& \hspace{-1cm} \frac{d^{2}E_{2_{b}}}{d\omega d\Omega }=\frac{n\omega^{2}}{4\pi ^{2}}\frac{\tilde{\theta}^{2}q^{2}}{4n^{2}
+\tilde{\theta}^{2}}\left( 1-\frac{1}{v^{2}n^{2}}\right)\frac{\sin^{4}\left( \frac{\zeta }{2}\tilde{\Xi}_{-} \right) }{\frac{1}{4}\tilde{\Xi}_{-}^{2} }. \label{d2E 2 Im explicit}
\end{eqnarray}%
The presence of $|\cos \theta| $ in $\tilde{\Xi}_{-}$ indicates that now the radiation is concentrated in two main lobes, one around $\theta_C$, contributing to the forward radiation, and the other around $\pi - \theta_C$, contributing to the reversed VC radiation. Taking into account both lobes when integrating Eq.(\ref{d2E 2 Re explicit}) over the solid angle and  using Eq. (\ref{LOBE_EST}), we find 
\begin{equation}
\frac{dE_{2_{a}}}{d\omega}=\frac{\tilde{\theta}^{2}q^{2}}{4n^{2}+\tilde{\theta}^{2}}\frac{\omega L}{{2}}\left( 1-\frac{1}{v^{2}n^{2}}\right) .
\label{dE 2 Re}
\end{equation}
The term  ${d^2E_{2b}}/{d \omega d \Omega}$ in Eq. (\ref{d2E 2 Im explicit}) also contributes both to the forward and backward radiation and differs from the case of (\ref{d2E 2 Re explicit}) by the replacements $\tilde{\Xi}_{-}\rightarrow \tilde{\Xi}_{-}/2$ and $\sin^2 u \rightarrow \sin^4 u$. Nevertheless, we obtain the same result as in Eq. (\ref{dE 2 Re}),
\begin{equation}
\frac{dE_{2_{b}}}{d\omega}=\frac{\tilde{\theta}^{2}q^{2}}{4n^{2}+\tilde{\theta}^{2}}\frac{\omega L}{2}\left( 1-\frac{1}{v^{2}n^{2}}\right).
\label{dE 2 Im}
\end{equation}%
Therefore, the total radiated  energy per unit frequency is 
\begin{eqnarray}  \label{dE dk0}
\frac{dE}{d\omega}&=& \frac{dE_1}{d\omega}+\frac{dE_{12}}{d\omega}+\frac{dE_{2a}}{d\omega}+\frac{dE_{2b}}{d\omega},\nonumber\\
& =&q^{2}\omega L \left(1-\frac{1}{v^{2}n^{2}}\right),
\label{FULLCR}
\end{eqnarray}
which corresponds to the same expression as in the absence of the $\vartheta$ medium. At this stage, we do not know whether  Eq. (\ref{FULLCR}) is just a coincidence or there is a fundamental reason for this result. For our purposes, the main point is that the  total energy distribution (\ref{FULLCR}) is split into a nonzero reversed VC radiation (RVCR) 
\begin{eqnarray}
\frac{dE_{\rm RVCR}}{d\omega}&=&\frac{1}{2}\left(\frac{dE_{2a}}{d\omega}+\frac{dE_{2b}}{d\omega} \right), \nonumber \\
&=& q^{2}\omega L \left(1-\frac{1}{v^{2}n^{2}}\right)\Bigg[ \frac{1}{2}\, \frac{{\tilde \theta}^2}{4n^2+ {\tilde \theta}^2}\Bigg],
\label{RCR}
\end{eqnarray}
together with  a modified forward VC radiation  (FVCR)
\begin{equation}
\frac{dE_{\rm FVCR}}{d\omega}=q^{2}\omega L \left(1-\frac{1}{v^{2}n^{2}}\right)\Bigg[1- \frac{1}{2} \, \frac{{\tilde \theta}^2}{4n^2+ {\tilde \theta}^2}\Bigg].
\label{FCR}
\end{equation}  
We can restate expression (\ref{RCR}) in terms of the  number of photons radiated per unit length and per unit frequency as 
\begin{equation}  \label{dN dz}
\frac{d^2N_{\rm RVCR}}{d L d\omega }=\alpha \left(1-\frac{1}{v^{2}n^{2}}\right)%
\left[\frac{1}{2}\, \frac{\tilde{\theta}^{2}}{4n^{2}+\tilde{\theta}^{2}%
}\right].
\end{equation}
Another relevant parameter is the  power radiated per unit frequency in the backward direction,
\begin{equation}  
\frac{d^2E_{\rm RVCR}}{d t d\omega }={v}\frac{d^2E_{\rm RVCR}}{ dL d \omega }=q^{2}\omega v\left(1-\frac{1}{v^{2}n^{2}}\right)\Bigg[ \frac{1}{2}\, \frac{{\tilde \theta}^2}{4n^2+ {\tilde \theta}^2}\Bigg],
\label{d2E dt dw}
\end{equation}
whose estimation in Table \ref{TABLA1} will be used  for further comparison with alternative predictions.

\section{Order of magnitude  evaluations}
\label{NUM_EST}
In order to present some numerical estimations we use the setup in Fig. {\ref{REGIONS}} where medium 1 is a regular insulator with $\epsilon=4, \mu=1, \vartheta=0$, and  medium 2 is the topological insulator  ${\rm TlBiSe}_2 $ with $\epsilon=4, \, n=2, \, \mu=1$,  ${\vartheta}= \pi$,  and  $0 < m < 5 $ \cite{TlBiSe2}. This choice of $m$  yields a value of $\tilde \theta$ in the range $[\alpha, 11\alpha]$, which very much suppresses the backward radiation as compared with  the forward output.  Our  numerical estimations do not consider the frequency dependence of the refraction index.
In this way,  we can calculate the ratio
\begin{equation}
\frac{dE_{\rm RVCR}/{d\omega }}{dE_{\rm FVCR}/{ d\omega }}=\frac{1}{8} \left(\frac{{\tilde \theta}}{n}\right)^2,
\label{RATIO}
\end{equation}
where we also take  $4n^2 \gg {\tilde \theta}^2$. 
For the topological insulator  ${\rm TlBiSe}_2 $  the above ratio ranges between   $1.7 \times 10^{-6}$ and $2.0 \times 10^{-4}$ for the choices ${\tilde \theta}=\alpha$  and ${\tilde \theta}=11 \alpha$, respectively. In Table \ref{TABLA1}, we show the  power radiated per unit frequency in the backward direction, given by  Eq. (\ref{d2E dt dw}), for ${\rm TlBiSe}_2 $ with $\tilde \theta=11 \alpha$. The results are  in units of $\left[{\mu W}/{eV}\right]$, for $\omega\in[2,\,8]$ eV.  This range includes  the major sector  of the frequency spectrum  in the  forward VC radiation. The conversion to  MKS units  is 1$\,\mu$W/eV= 6.24$\times$ 10$^{12}$ s$^{-1}$, which  is  equal to $4.11\times 10^{-4}$ eV
 in units where $c=\hbar=1$.

\section{Summary}
\label{Summary}

We have considered the radiation produced by an electric charge propagating with constant velocity $v$ in the direction  ${\mathbf {\hat u}}$ (shown in  Fig. \ref{REGIONS})   between  two $\vartheta$ media with the same permittivity,   
whose electromagnetic response is driven by the modified Maxwell equations (\ref{HOMEQ})--(\ref{TILDE_THETA}). 
When $v$ is higher than the speed of light in the media, we discover the emission of reversed 
VC radiation, codified in  the angular distribution given by  Eq. (\ref{dE 2 explicit}) and illustrated in Figs. \ref{CHE2}--\ref{FULLCHE}.
These right-handed $\vartheta$ media are realized in nature as  magnetoelectric materials, among which we find  topological insulators, having positive permittivity, permeability, and index of refraction. 

The main characteristics of the reversed VC radiation we have discovered  are the following: (i) The threshold condition $v>c/n$
for the velocity of the charge must be satisfied as in the standard case. (ii) The reversed VC radiation occurs for all  frequencies in the VC spectrum and it is always accompanied by  forward VC radiation. (iii) The energy loss per unit frequency  of the reversed  VC radiation is highly  suppressed  with respect to the forward output according to Eq. (\ref{RATIO}). 
 A comparison with  measurements of reversed VC radiation in metamaterials can be established   by 
interpreting this suppression as due to the detection of radiation at an effective frequency $\omega_{\rm eff}=\omega \, {\tilde \theta}^2/8 n^2$, according to Eq. (\ref{RCR}). Standard \v{C}erenkov counters work in the range of $140-800\, {\rm nm}$ corresponding to 
$8.9-1.6  \,\, {\rm eV}$, respectively. 
Taking an average of $500 \,\, {\rm nm}$ ($2.5 \,\, {\rm eV}$), we would expect detectable reversed VC radiation  at  $\omega_{\rm eff}$ in the range from $4\times 10^{-3} \, {\rm meV}$ for ${\tilde \theta}=\alpha$ to $0.5 \;{\rm meV} $  for ${\tilde \theta}=11 \alpha$,  respectively, using ${\rm TlBiSe}_2$  as a  $\vartheta$ medium. However, recent measurements of reversed VC radiation in metamaterials  show that these estimations are within the experimental capabilities. Reversed VC radiation has been measured at a frequency of $2.85 \,\, {\rm  GHz}$, equivalent to $1.2 \times 10^{-2} \,\, {\rm meV}$, in an all-metal metamaterial consisting of a square waveguide loaded with complementary electric split-ring resonators \cite{Duan}. Likewise, reversed VC radiation in the range $(3.4-3.9) \times 10^{-2}\,\, {\rm meV} $ has also been experimentally verified in  a phased electromagnetic dipole array used to model a moving charged particle \cite{Sheng}.

Our estimations for $d^2 E_{\rm RVCR}/dt d\omega$ in Table \ref{TABLA1} are in the range $10^{-3}-10^{-2} \,\,  \mu{\rm W/ eV}$,  in the frequency interval of  $2-8 \,\, {\rm eV}$. 
They are  smaller by a factor of $10^{-4}-10^{-3}$  than the maximum output of $\sim 10 \,  \mu{\rm W/ eV}$ theoretically predicted to occur in the narrow  interval of $5.7-6.5$ eV in a metal-insulator-metal waveguide \cite{Tao}.  In such a waveguide with a core thickness of  $a=20 \,\, {\rm nm}$, surface plasmon polaritons excited by an electron moving at $v=0.8$ produce reversed VC radiation.  

A qualitative argument for the existence of the reversed \v{C}erenkov radiation in a $\vartheta$ media can be given by extending the interpretation of the static fields produced by a point charge located in front of a topological insulator in terms of electric and magnetic images \cite{UrrutiaMartinCambiaso, QI_SCIENCE}.  When the charge $q$ moves from region {1} to region {2}, as we have  assumed here, the effect of the interface $\Sigma$  can be replaced  by  introducing a moving image electric charge ${\tilde q}=-q {\tilde \theta}^2/(4 n^2+{\tilde \theta}^2)$  together with a moving  image magnetic monopole ${\tilde g}=2q {\tilde \theta}/(4 n^2+{\tilde \theta}^2)$, both located in region {2} and moving towards region {1} . These images would contribute to the physical fields only in region {1} with their own forward VC radiation in the  far-field zone, which turns out to be in the  reversed direction  with respect to the incident charge. As shown in Ref. \cite{Schwinger}, a magnetic monopole $g$ propagating with a velocity $v> 1/n$  also produces  a  forward VC cone, with  radiated energy  per unit length and per unit frequency  given by
\begin{equation}
\frac{d ^2E_{\rm monopole}}{dL d\omega }={\omega g^2 n^2}\left(1-\frac{1}{v^2 n^2}\right),
\end{equation}
in complete analogy with Eq. (\ref{DEDOMEGAST}). From this point of view, the dominant contribution to the reversed VC radiation of the incident charge  would arise from the image magnetic monopole since $g \sim {\tilde \theta}$ and the image electric charge would contribute with higher order terms of order  ${\tilde q}^2 \sim {\tilde \theta}^4$. A detailed calculation of the radiation fields in region 1  produced by  this configuration is further required to test this interpretation, in particular, to verify that the final factor depending on $\tilde \theta $ will have the correct form given in Eq. (\ref{RCR}).  As it is well known, image charges are only useful mathematical tools in the description of electromagnetic phenomena and do not represent physically existing entities. In our case, the physical response of the medium is produced by  the time-varying electric  charge densities and  Hall currents induced at the interface $\Sigma$  due to the magnetoelectric effect. The verification of the above  interpretation of the reversed VC radiation is beyond the scope of the present paper and  is postponed for future work.

\begin{table}
\begin{center}
\begin{tabular}{cc}
\hline 
\hline
{{} $\omega \,\, ({\rm eV)}$ } \qquad   & {\quad {}$  {d^{2}E_{{\rm RVCR}}}/{dtd\omega}$ $\left({\mu W}/{{\rm eV}}\right)  $}  \tabularnewline
\hline 
\hline
$2$ \qquad & \qquad $3.5 \times  10^{-3}$ \quad \tabularnewline
$4$ \qquad & \qquad $7.0 \times  10^{-3}$ \quad  \tabularnewline
$6$ \qquad & \qquad $1.0 \times  10^{-2}$ \quad \tabularnewline
$8$\qquad & \qquad $1.4 \times  10^{-2}$  \quad  \tabularnewline
\hline
\hline 
\end{tabular}
\caption{This table shows the orders of magnitude for the power radiated per  unit frequency in the backward direction (reversed VC radiation) when a particle with $v=0.8$ and charge  $q=\sqrt{\alpha}$ propagates  across the interface of  a normal insulator and the  topological insulator ${\rm TlBiSe}_2 $ characterized by  $n=2$, $\tilde{\theta}=11\alpha$.}
\par
\label{TABLA1}
\end{center}
\end{table}

Our results apply for any material  that supports the magnetoelectric  effect,  in such a way that its electromagnetic response can be described by $\vartheta$-ED. In the particular case of a topological insulator, the conditions for the realization of such  effect  are (i) the topological insulator should be in the 3D regime, (ii) all the surfaces need to be  gapped with the chemical potential lying within the gaps and (iii) the dynamics of the  interior of the topological insulator should be invariant under time-reversal symmetry or inversion symmetry, in order  to keep $\vartheta=\pi$ in the bulk \cite{DiXiao}. In this work, we have highlighted the case of  3D strong topological insulators, which require time-reversal symmetry breaking in the surfaces to realize the topological magnetoelectric effect. This can be achieved by magnetically gapping  all  the interfaces, in such a way that the  entire sample behaves as an insulator having a magnetoelectric coupling of exactly 
$\vartheta=\pi$ \cite{VDB5}.  Since we have assumed that charged particles are moving with constant velocity across the material, it would  be advisable to break the time-reversal symmetry by doping  the surfaces with thin ferromagnetic films instead of switching on an external magnetic field. In any case, the velocity of such moving  particles has the lower limit of $c/n$ but must be large enough so that there is no appreciable deflection of the charge when going through the magnetically doped material. The  velocity of the external  moving charges is independent of the Fermi velocity of the electrons on the 2D surfaces and the Fermi energy can comfortably lie in the middle of the gap between the Dirac cones. We expect our results to hold also in the case of magnetically doped AXIs. They are heterostructures in which magnetic ions are added to the vicinity of the top and bottom interfaces of a 3D strong topological insulator, like (Bi,Sb)$_2$Te$_3$, for example. In this way, their  electromagnetic response is still coded in $\vartheta$-ED. Even though the   upward-downward  magnetic coating  in opposite interfaces will  produce a change in the sign of $\vartheta=\pi$ when going from one interface  to the other, this will  make the  contribution of both  interfaces to the spectral distribution of the reversed VC  radiation to add up. This is because each contribution depends  on ${\tilde \theta}^2$, respectively, according to Eqs. (\ref{TILDE_THETA_1}) and (\ref{dE 2 explicit}).

Finally, we comment on the condition $\epsilon_1= \epsilon_2$, imposed at the beginning of Sec. \ref{theta-ED}  
in order to get rid of the otherwise  unavoidable transition radiation and aimed to present a  clean derivation of the reversed VC radiation.   We do not mean that relaxing the equal permittivity condition  will eliminate the presence of reversed VC radiation, but we only claim that the choice $\epsilon_1 $ very different from $\epsilon_2$ will make transition radiation   interfere with the VC radiation  requiring a new  theoretical discussion as well as hindering any experimental detection. In fact, the spectral distribution of the transition radiation might also display unexpected  corrections  arising from  $\vartheta$-ED. In realistic terms,  the equal permittivity condition   
means  that  $\epsilon_1$ and  $ \epsilon_2$  could  be chosen such that their values are as close as possible.  This condition can be considered as a useful restriction in the planning of a possible experimental setup and looks plausible  since  the refraction indices  of normal insulators cover a wide range of values.

\acknowledgments

O. J. F. has been supported by the doctoral fellowship CONACYT-271523. O. J. F., L. F. U.  and O. R. T. acknowledge support from the CONACYT Project No. 237503. Support from the project No.  IN103319 from Direcci\'on General de Asuntos del
Personal Acad\'emico (Universidad Nacional Aut\'onoma de M\'exico) is also ackowledged. L. F. U.  and O. J. F. thank Dr. Alberto Mart\'\i n-Ruiz, Professor Hugo Morales-T\'ecotl and Professor Rub\'en Barrera for useful discussions and suggestions. The authors also thank Dr. Christine Gruber for a careful reading of the manuscript together with many helpful comments.

\appendix

\section{Green's function in the radiation zone}

\label{GFfarzone}

In  this Appendix, we obtain the  far-field approximation  for the GFs in Eq. (\ref{GF oscillating}) 
through the stationary phase method \cite{Chew}. 
We review the contribution 
of standard ED and consider only ${\bar G}^{\mu}_{\tilde{\theta}\;\nu}(\mathbf{x},\mathbf{x}^{\prime};\omega)$ 
to  illustrate the procedure of how to deal with the additional  contributions arising from  ${\cal L}_\vartheta$. 

\begin{figure}[tbp]
\centering 
\includegraphics[width=6CM]{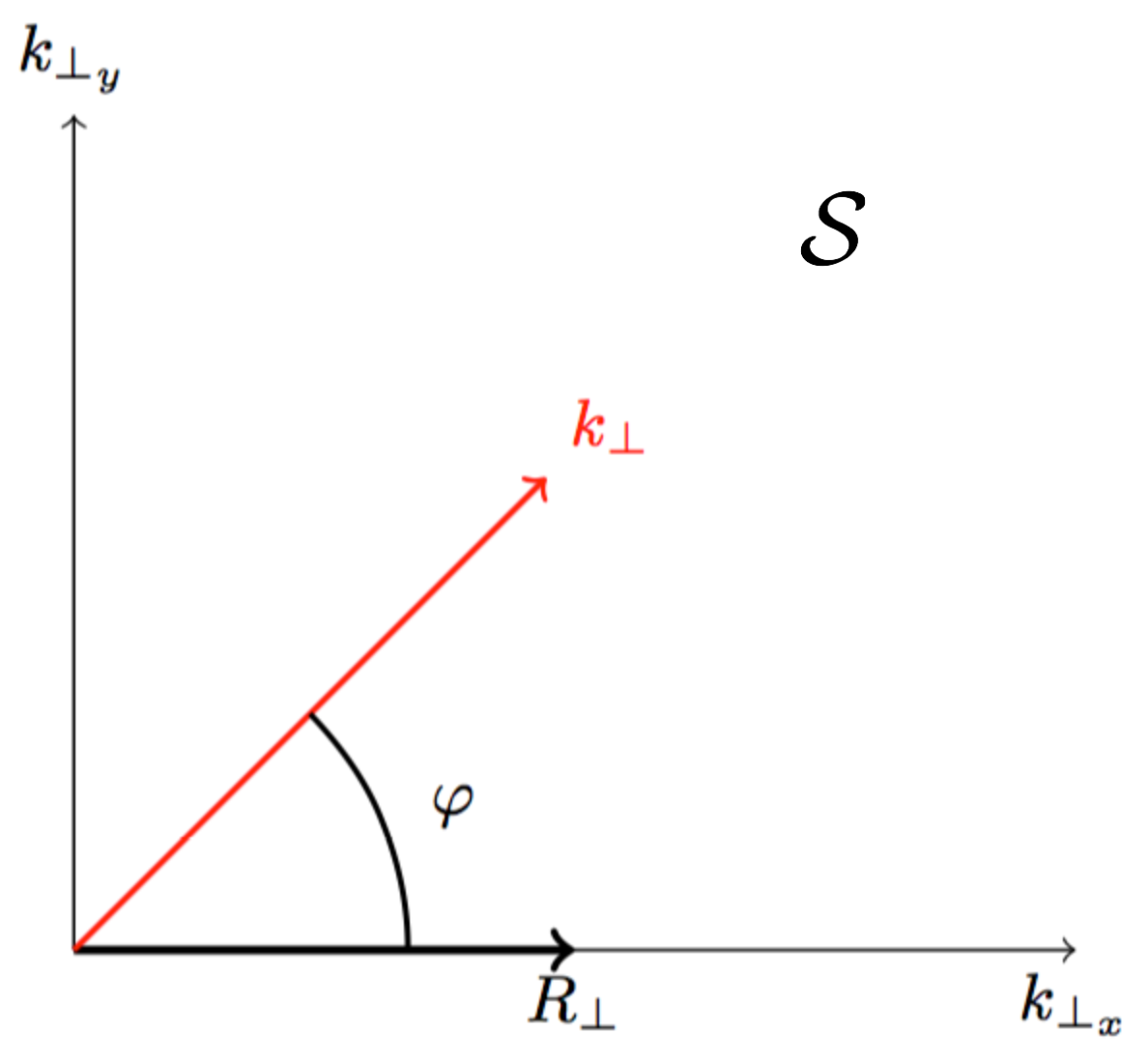} 
\caption{The vectors $\mathbf{k_{\perp}}, \mathbf{%
k_{\perp_{x}}}, \mathbf{k_{\perp_{y}}}$, and  $\mathbf{R_{\perp}}$  in the coordinate system ${\cal S}$.}
\label{SYSTEM}
\end{figure}

Let us start  with the components 
${\bar G}_{ED\;\nu }^{\mu }(\mathbf{x},\mathbf{x}^{\prime};\omega )$. 
The double integral in $\mathbf{k}_{\perp }$ is conveniently calculated by   expressing the area
element in polar coordinates $d^{2}\mathbf{k_{\perp}}=k_{\perp}dk_{\perp}d\varphi$ and choosing 
the $k_{\perp x}$ axis in the direction of the vector 
$\mathbf{R}_{\perp}=(\mathbf{x}-\mathbf{x^{\prime}})_{\perp}$. This defines the  coordinate 
system ${\cal S}$ to be repeatedly used in the following  and shown in the Fig. \ref{SYSTEM}.  Writing  $\mathbf{k_{\perp}}\cdot\mathbf{R}%
_{\perp}=k_{\perp}R_{\perp}\cos\varphi$ and recalling that the angular integral provides 
a representation of the Bessel function $J_{0}(k_{\perp}R_{\perp})$ \cite{Abramowitz} we obtain

\begin{eqnarray}
&&\hspace{-.5cm}{\bar G}^{\mu}_{ED\;\nu}(\mathbf{x},\mathbf{x}^{\prime};\omega)=i\eta^{\mu}_{\;\;%
\nu}\int_{0}^{\infty}\frac{k_{\perp}dk_{\perp}}{\sqrt{\tilde{k}_{0}^{2}-k_{\perp}^{2}%
}}J_{0}(k_{\perp}R_{\perp})\nonumber\\
&&\hspace{2cm}\times e^{i\sqrt{\tilde{k}_{0}^{2}-k_{\perp}^{2}}%
|z-z^{\prime}|}=\eta^{\mu}_{\;\;%
\nu} \frac{e^{i\tilde{k}_{0} R}}{R},
\label{SOMM_ID}
\end{eqnarray}
with  $R=\|\mathbf{x}-\mathbf{x}^{\prime}\|=\sqrt{R_{\perp}^{2}+(z-z^{%
\prime})^{2}}$, where  the right-hand side of the above equation is a direct consequence of the  Sommerfeld
identity \cite{Sommerfeld}.  Moreover, the coordinate conditions \begin{eqnarray}
\|\mathbf{x}\|\gg\|\mathbf{x}%
^{\prime}\|, && \,\,\,   R_{\perp}=\|\left(\mathbf{x}-\mathbf{x}%
^{\prime}\right)_{\perp}\|\simeq\|\mathbf{x}_{\perp}\|=\rho, \nonumber \\
&& |z|+|z^{\prime}|\simeq|z|\;,
\label{FFC}
\end{eqnarray}
in the far-field approximation produce the  
 well-known result for  ${\bar G}^{\mu}_{ED\;\nu}$ in standard ED \cite{Schwinger},
\begin{equation}  \label{G mu nu far zone ED}
{\bar G}^{\mu}_{ED\;\nu}(\mathbf{x},\mathbf{x}^{\prime};\omega)\rightarrow \eta^{\mu}_{\;\;\nu}%
\frac{e^{i\tilde{k}_{0}(r-\hat{\mathbf{n}}\cdot\mathbf{x}^{\prime})}}{r},
\end{equation}
with $\hat{\mathbf{n}}$ being a unit vector in the direction of $\mathbf{x}$ and $\|\mathbf{x}\|=r$ .

For future purposes, it will be convenient to  go back to the generalized  Sommerfeld identity in Eq.(\ref{SOMM_ID0}) and  to rewrite it in terms of Hankel functions. We start from   
\begin{equation}
J_{0}(x)=\frac{1}{2}\left(H_{0}^{(1)}
(x)+H_{0}^{(2)}(x)\right),
\end{equation}
where $H_{0}^{(1)}$ and $H_{0}^{(2)}$ are the Hankel functions, together with 
the reflection formula
$H_{0}^{(1)}(e^{i\pi}x)=-H_{0}^{(2)}(x)$, which allows us to extend the integration interval to $-\infty$. The result is 
\begin{equation}  \label{Sommerfeld 2}
\frac{e^{i\tilde{k}_{0} {\cal R}}}{{\cal R}}=\frac{i}{2}\int_{-\infty, C}^{\infty}\frac{%
k_{\perp}dk_{\perp}}{\sqrt{\tilde{k}_{0}^{2}-k_{\perp}^{2}}}H_{0}^{(1)}(k_{\perp}R_{%
\perp})e^{i\sqrt{\tilde{k}_{0}^{2}-k_{\perp}^{2}}|Z|},
\end{equation}
with ${\cal R}=\sqrt{R_\perp^2+ Z^2}$ and where  $Z$ can be conveniently chosen. Here $C$ denotes the path of integration  defined in Fig. 2.2.5
of Ref. \cite{Chew}. 

Now, we calculate the far-field approximation of   ${\bar G}^{\mu}_{\tilde{\theta}\;\nu}$. Starting from the second expression in  Eqs. (\ref{GF oscillating}) and  
choosing the  coordinate system ${\cal S}$ previously introduced in  Fig. \ref{SYSTEM}, ${\bar G}^{\mu}_{\tilde{\theta}\;\nu}$ can be rewritten as 
\begin{equation}  \label{G mu vu theta 1 I alpha}
{\bar G}^{\mu}_{\tilde{\theta}\;\nu}(\mathbf{x},\mathbf{x}^{\prime};\omega)=i%
\varepsilon^{\mu\;\;\alpha 3}_{\;\;\nu}\frac{2\tilde{\theta}}{4+\tilde{\theta%
}^{2}}I_{\alpha}(\mathbf{x},\mathbf{x}^{\prime};\omega),
\end{equation}
where we define the following  integrals:
\begin{eqnarray}  \label{I alpha}
I_{\alpha}(\mathbf{x},\mathbf{x}^{\prime};\omega)&=&\int_{0}^{\infty}\frac{%
k_{\perp}dk_{\perp}}{\tilde{k}_{0}^{2}-k_{\perp}^{2}}e^{i\sqrt{\tilde{k}_{0}^{2}-k_{%
\perp}^{2}}(|z|+|z^{\prime}|)}\nonumber\\
&&\times\int_{0}^{2\pi}\frac{d\varphi}{2\pi}%
e^{ik_{\perp}R_{\perp}\cos\varphi}k_{\alpha},
\end{eqnarray}
which we subsequently separate according to $\alpha=0$ and $ \alpha= k=1,2$. Rewriting the angular integration in terms of Bessel functions and employing the expression (\ref{Sommerfeld 2}) for the Sommerfeld integral with $Z=|z|+|z'|$, we obtain
\begin{eqnarray}
I_{0}(\mathbf{x},\mathbf{x}^{\prime};\omega)&=&\frac{\omega}{2}
\int_{-\infty, C}^{\infty}\frac{k_{\perp}dk_{\perp}}{\tilde{k}_{0}^{2}-k_{\perp}^{2}
}H_{0}^{(1)}(k_{\perp}R_{\perp})\nonumber\\
&&\times e^{i\sqrt{\tilde{k}_{0}^{2}-k_{\perp}^{2}}
(|z|+|z^{\prime}|)},  \label{I0 stationary phase 1} \\
{I_k}(\mathbf{x},\mathbf{x}^{\prime};\omega)&=&\frac{i}{2R_{\perp}} (\mathbf{x}-\mathbf{x'})_{\perp k}
\frac{\partial}{\partial R_{\perp}}\int_{-\infty, C}^{\infty}\frac{%
k_{\perp}dk_{\perp}}{\tilde{k}_{0}^{2}-k_{\perp}^{2}}\nonumber\\
&&\times H_{0}^{(1)}\left(k_{\perp}R_{%
\perp}\right)e^{i\sqrt{\tilde{k}_{0}^{2}-k_{\perp}^{2}}(|z|+|z^{\prime}|)}.
\label{Ii stationary phase 1}
\end{eqnarray}

Next, we implement the far-field conditions (\ref{FFC}),
yielding   
\begin{equation}
H_{0}^{(1)}\left(k_{\perp}\rho\right)\sim\sqrt{\frac{2}{\pi k_{\perp}\rho}}%
e^{ik_{\perp}\rho-i\frac{\pi}{4}}
\end{equation}
in the limit $%
\rho\rightarrow\infty$ and $z\rightarrow\infty$. Then, the following product reduces to  
\begin{equation}
H_{0}^{(1)}\left(k_{\perp}\rho\right)e^{i\sqrt{\tilde{k}_{0}^{2}-k_{\perp}^{2}}%
|z|}\sim\sqrt{\frac{2}{\pi k_{\perp}\rho}}e^{ik_{\perp}\rho-i\frac{\pi}{4}+i%
\sqrt{\tilde{k}_{0}^{2}-k_{\perp}^{2}}|z|},
\label{STA_PHASE}
\end{equation}
in each of the integrands of  Eqs. (\ref{I0 stationary phase 1}) and (\ref{Ii stationary phase 1}). 
This factor is a rapidly oscillating function of $k_{\perp}$ that allows one to apply the stationary phase approximation  to evaluate the dominant contribution \cite{Chew,Chew1}. Recalling that $k_{z}=\sqrt{\tilde{k}_{0}^{2}-k_{\perp}^{2}}$, the stationary phase condition  is 
\begin{equation}
(k_{\perp})_s=\tilde{k}_{0}\rho/r, \quad  (k_z)_s=\tilde{k}_{0}|z|/r,
\label{ST_PHASE_COND}
\end{equation} 
with $r=\sqrt{\rho^{2}+z^{2}}$. Moreover, in Eqs. (\ref{I0 stationary phase 1}) and (\ref{Ii stationary phase 1}) we can  estimate  $\tilde{k}_{0}^{2}-k_{\perp}^{2}=k_{z}^{2}\simeq (k_z)_s k_{z}$ around the
 point where the phase is stationary. Using again  the Sommerfeld identity (\ref{Sommerfeld 2}) with ${\cal R}={\tilde R}$ and $Z=|z|+|z'|$,
we arrive at 
\begin{eqnarray}
&& \hspace{-1cm}I_{0}(\mathbf{x},\mathbf{x}^{\prime};\omega)=\frac{\omega r}{i\tilde{k}_{0}|z|}\frac{e^{i\tilde{k}_{0}\tilde{R}}}{\tilde{R}},
\label{I0 stationary phase 2} \\
&&\hspace{-1cm} {I_k}(\mathbf{x},\mathbf{x}^{\prime};\omega)
=i \frac{(\mathbf{x}-\mathbf{x'})_{\perp k}}{\tilde{k}_{0}|z|}\, r\, \left[\frac{\tilde{k}_{0}}{\tilde{R}^{2}}+\frac{i}{\tilde{R}^{3}}\right]e^{i\tilde{k}_{0}\tilde{R}},
\label{Ii stationary phase 2}
\end{eqnarray}
where 
\begin{equation}
\tilde{R}=\sqrt{\left(\mathbf{x}-\mathbf{x}^{\prime}\right)_{%
\perp}^{2}+(|z|+|z^{\prime}|)^{2}}.
\label{RTILDE1}
\end{equation}
 Finally, we complete  the far-field  approximation in Eqs. (\ref{I0 stationary
phase 2}) and (\ref{Ii stationary phase 2}) by writing
\begin{equation}
\tilde{R}=r-\mathbf{n}_{\perp}\cdot\mathbf{x}^{\prime}_{\perp}+|z^{\prime}\cos\theta|,
\label{RTILDE}
\end{equation}
with $\mathbf{n}_\perp = \sin\theta ( \cos \phi, \, \sin \phi, \, 0)$. The results are
\begin{eqnarray}
&&  \hspace{-0.8cm} I_{0}(\mathbf{x},\mathbf{x}^{\prime};\omega)=\frac{e^{i\tilde{k}_{0} r}}{inr|\cos\theta|}e^{i\tilde{k}_{0}\left(-\mathbf{n}_{\perp}\cdot\mathbf{x}%
^{\prime}_{\perp}+|z^{\prime}\cos\theta|\right)},  \label{I0 final} \\
&& \hspace{-0.8cm} I_{k}(\mathbf{x},\mathbf{x}^{\prime};\omega)=\frac{{n}_{\perp k} \, 
 e^{i\tilde{k}_{0} r}}{ir|\cos\theta|}e^{i\tilde{k}_{0}\left(-\mathbf{n}%
_{\perp}\cdot\mathbf{x}^{\prime}_{\perp}+|z^{\prime}\cos\theta|\right)}.
\label{Ii final}
\end{eqnarray}
where we have dropped terms of $\mathcal{O}(r^{-2})$ and higher.

Next we comment on a  technical point,  which fortunately is not relevant to our purposes. On one hand, the integrand in Eq. (\ref{I0 stationary phase 1}) has  poles at $k_\perp=\pm {\tilde k}_0$.  At the same time, the stationary point obtained in Eq. (\ref{ST_PHASE_COND}) indicates that $(k_\perp)_s={\tilde k}_0 \sin \theta$. Both values coincide for $\theta=\pm \pi/2$ indicating the factor of the exponential is not a smooth function around the stationary point now. In this way, the stationary phase calculation  requires some modifications, which we do not pursue in this work. The reason  we do not require such improvements is because the VC cone  condition demands $0<{\theta}_{\rm C} < \cos^{-1}(1/n)$, which is always far away from  the dangerous point $\theta=\pi/2$,  unless $n$ is extremely large. In other words, our calculation in  the stationary phase method is still a good approximation for the far-field behavior of our GFs when $\theta < \pi/2$.  An analogous situation occurs when dealing with the steepest
descent method \cite{WAIT, VANDER, CLEMM}.

At this point and
as a matter of consistency, we show that the stationary phase method and the steepest
descent method lead to the same results, under a similar restriction for ${\theta}_{\rm C}$. Let us recall that the latter  method \cite{Chew, Mandel} allows the calculation of the leading  contribution to the integral
\begin{equation}
I=\int e^{\lambda h(t)}f(t)dt,
\label{INT1}
\end{equation}
when  a rapidly varying exponential factor  multiplies the function $f(t)$, which must have  a smooth behavior in the region close to the stationary phase point, yielding the result 
\begin{equation}
I\sim e^{\lambda h(t_{0})}f(t_{0})\sqrt{\frac{-2\pi}{\lambda h''(t_{0})}},
\label{SADDLE_APROX}
\end{equation}
where $t_0$ is such that $h'(t_0)=0$. 
From Eqs. (\ref{I0 stationary phase 1}) and (\ref{Ii stationary phase 1}), we extract the 
required integral as 
\begin{equation}  \label{K1 saddle point}
K_{1}(\mathbf{x};\omega)=\int_{-\infty, C}^{\infty}\frac{k_{\perp}dk_{\perp}%
}{\tilde{k}_{0}^{2}-k_{\perp}^{2}}H_{0}^{(1)}(k_{\perp}R_{\perp})e^{i\sqrt{%
\tilde{k}_{0}^{2}-k_{\perp}^{2}}(|z|+|z^{\prime}|)}\;,
\end{equation}
whose asymptotic behavior for $\tilde{k}_{0}^{2}>k_{\perp}^{2}$ is
\begin{equation}
K_{1}(\mathbf{x};\omega)\sim\int_{-\infty, C}^{\infty}\frac{%
k_{\perp}dk_{\perp}}{\tilde{k}_{0}^{2}-k_{\perp}^{2}}\sqrt{\frac{2}{\pi k_{\perp}\rho%
}}e^{ik_{\perp}\rho-i\frac{\pi}{4}}e^{i\sqrt{\tilde{k}_{0}^{2}-k_{\perp}^{2}}%
|z|}.
\label{K1}
\end{equation}
Since the rapidly oscillating phase in Eq. (\ref{K1}) coincides with that of Eq. (\ref{STA_PHASE}) in the far-field regime, the stationary point of $K_{1}(\mathbf{x};\omega)$ is also given by the conditions in Eq. (\ref{ST_PHASE_COND}), yielding
\begin{equation}  \label{saddle phase}
e^{ik_{\perp}R_{\perp}-i\frac{\pi}{4}}e^{i\sqrt{\tilde{k}_{0}^{2}-k_{\perp}^{2}}%
(|z|+|z^{\prime}|)}\big|_{k_{\perp}=k_{\perp_{s}}}=e^{i\tilde{k}_{0}\tilde{R}-i\frac{%
\pi}{4}}
\end{equation}
as the  phase at the stationary point. 
After this, we require computation of the second derivative of the phase at the
stationary point in the far-field regime, which is 
\begin{equation}  \label{saddle second derivative}
\frac{d^{2}}{dk_{\perp}^{2}}\left(k_{\perp}\rho-\frac{\pi}{4}+\sqrt{%
\tilde{k}_{0}^{2}-k_{\perp}^{2}}|z|\right)\bigg|_{k_{\perp}=k_{\perp_{s}}}=-\frac{%
r^{3}}{\tilde{k}_{0}z^{2}}\;.
\end{equation}
Substituting Eqs. (\ref{saddle phase}) and (\ref{saddle second derivative}) in
Eq.  (\ref{SADDLE_APROX}), we obtain that the leading-order term calculated  by 
 this method is 
\begin{equation}
K_{1}(\mathbf{x};\omega)\sim e^{i\tilde{k}_{0}(r-\mathbf{n}_{\perp}\cdot\mathbf{x}^{\prime}_{\perp}+|z^{%
\prime}\cos\theta|)}\frac{2}{i\tilde{k}_{0}|z|},
\label{K11}
\end{equation}
where we have disregarded ${\pi}/{4}$ in front of  $%
\tilde{k}_{0}r \rightarrow \infty$ and we have replaced ${\tilde R}$ by its expression (\ref{RTILDE}). Inserting Eq. (\ref{K11}) in the expressions (\ref{I0 stationary phase 1}) and 
(\ref{Ii stationary phase 1}), we obtain the results given in Eqs. (\ref{I0
final}) and (\ref{Ii final}) by the stationary phase method.
Therefore, putting together Eqs. (\ref{I0 final}) and (\ref{Ii final}) in Eq. (\ref{G mu vu theta 1 I alpha}), we find the final expression for ${\bar G}^{\mu}_{%
\tilde{\theta}\;\nu}(\mathbf{x},\mathbf{x}^{\prime};\omega)$  as
\begin{eqnarray}
{\bar G}^{\mu}_{\tilde{\theta}\;\nu}(\mathbf{x},\mathbf{x}^{\prime};\omega)&=&%
\varepsilon^{\mu\;\;\alpha3}_{\;\;\nu}\frac{2\tilde{\theta}}{4\epsilon+\tilde{\theta}%
^{2}}\frac{{s}_{\alpha}}{|\cos\theta|}\frac{e^{i\tilde{k}_{0} r}}{r}\nonumber\\
&&\times e^{i\tilde{k}_{0}\left(-\mathbf{n}_{\perp}\cdot\mathbf{x}^{\prime}_{\perp}+|z^{%
\prime}\cos\theta|\right)}\;,
\end{eqnarray}
where ${s}_{\alpha}=(1/n,\mathbf{\hat n})$. 

The calculation of 
$\bar{G}_{\tilde{\theta}^{2}\;\nu }^{\mu }$ proceeds along similar lines and the final result is summarized in the equations
\begin{eqnarray}
 \bar{G}_{\tilde{\theta}^{2}\;\nu }^{\mu }(\mathbf{x},\mathbf{x}^{\prime
};\omega )&=& \frac{\tilde{\theta}^{2}}{4n^{2} +\tilde{\theta}^{2}}\frac{%
e^{i\tilde{k}_{0}r}}{r\cos^{2}\theta}C^{\mu }_{\;\;\nu }(\mathbf{x},n)\nonumber\\
&&\times e^{i\tilde{k}_{0}\left( -\mathbf{n}_{\perp }\cdot 
\mathbf{x}_{\perp }^{\prime }+|z^{\prime }\cos \theta |\right)},
\label{GF far zone1} 
\end{eqnarray}
where

\begin{equation}
C^{\mu }_{\;\;\nu }(\mathbf{x},n)=\left( 
\begin{array}{cccc}
\sin ^{2}\theta \quad & -{x}/{rn}&-{y}/{rn}  & 0 \\ 
{x}/{rn}& -{1}/{n^{2}} & 0 & 0 \\ 
{y}/{rn}& 0 & -{1}/{n^{2}}+\sin^{2}\theta \quad  & 0 \\ 
0 & 0 & 0 & 0%
\end{array}%
\right).
\end{equation}
We have also verified that the steepest descent method \cite{Chew,Mandel} 
gives the same results as  the stationary phase approximation in the calculation of the required rapidly oscillating  integrals for this contribution.

\end{document}